\documentclass[useAMS,usenatbib]{mn2e}
\usepackage{amsmath}
\usepackage{amssymb}
\usepackage{graphicx}
\usepackage{epsfig}
\makeatletter{}
\def\Mpch{\mbox{$h^{-1}$Mpc }}

\def\Mvir{\mbox{$M_{\rm vir }$}}
\def\M200{\mbox{$M_{\rm 200 }$}}
\def\Msunh{\mbox{$h^{-1}M_\odot$ }}
\def\Rvir{\mbox{$R_{\rm vir }$}}
\def\R200{\mbox{$R_{\rm 200 }$}}
\def\Vcirc{\mbox{$V_{\rm circ }$}}
\def\Vmax{\mbox{$V_{\rm max }$ }}
\def\vmax{\mbox{$v_{\rm max }$ }}

\def\deltacr{\mbox{$\delta_{\rm cr}$ }}
\def\V200{\mbox{$V_{\rm 200 }$}}
\renewcommand{\vec}[1]{ {\bmath #1} }

\newcommand{\lsim}{\mbox{${\,\hbox{\hbox{$ < $}\kern -0.8em \lower 1.0ex\hbox{$\sim$}}\,}$}}
\newcommand{\gsim}{\mbox{${\,\hbox{\hbox{$ > $}\kern -0.8em \lower 1.0ex\hbox{$\sim$}}\,}$}}

\def\beqn{\vspace{2mm}
\begin{eqnarray}} 
\def\eeqn{\vspaceg{2mm} 
\end{eqnarray}}

\newcommand{\be}{\begin{equation}}
\newcommand{\ee}{\end{equation}}
\newcommand{\ba}{\begin{eqnarray}}
\newcommand{\ea}{\end{eqnarray}}
\newcommand{\brr}{\begin{array}}
 
\newcommand{\err}{\end{array}}
\newcommand{\bc}{\begin{center}}
\newcommand{\ec}{\end{center}}

\title[Structure of halos]{MultiDark simulations: the story of
  dark matter halo concentrations and density profiles.}

\makeatletter{}\author[A.~Klypin \& et.al.]
  {Anatoly~Klypin$^{1}$ and
  Et~Al$^{3,4}$ \vspace{0.2cm}\\ 
  $^1$Astronomy Department, New Mexico State University, Las Cruces, NM, USA\\
}

\author[Klypin et. al.]{Anatoly~Klypin$^1$\thanks{E-mail: aklypin@nmsu.edu},
Gustavo~Yepes$^2$, 
Stefan~Gottl{\"o}ber$^3$,
Francisco~Prada$^{4,5,6}$, 
\newauthor
and Steffen~He{\ss}$^3$ \\
 \vspace{-0.2cm}\\
$^{1}$	Astronomy Department, New Mexico State University, Las Cruces, NM, USA\\
$^{2}$	Departamento de F\' {\i}sica Te\'orica M8, Universidad Autonoma de Madrid (UAM), Cantoblanco, E-28049,  Madrid, Spain\\
$^{3}$	Leibniz-Institut f\"ur Astrophysik Potsdam (AIP), Potsdam, Germany\\
$^{4}$	 Instituto de F\'{\i}sica Te\'orica, (UAM/CSIC), Universidad Aut\'onoma
      de Madrid,  Cantoblanco, E-28049 Madrid, Spain \\
$^5$ Campus of International Excellence UAM+CSIC, Cantoblanco, E-28049 Madrid, Spain \\
$^6$ Instituto de Astrof\'{\i}sica de Andaluc\'{\i}a (CSIC), Glorieta de 
     la Astronom\'{\i}a, E-18080 Granada, Spain \\
\\
}

\begin{document}
\pagerange{\pageref{firstpage}--\pageref{lastpage}} 
\maketitle
\label{firstpage}
\begin{abstract}
Predicting structural properties of dark matter halos is one of the
fundamental goals of modern cosmology.  We use the suite of MultiDark
cosmological simulations to study the evolution of dark matter halo
density profiles, concentrations, and velocity anisotropies. We find
that in order to understand the structure of dark matter halos and to
make 1--2\% accurate predictions for density profiles, one needs to
realize that halo concentration is more complex than the ratio of the
virial radius to the core radius in the Navarro-Frenk-White profile.
For massive halos the average density profile is far from the NFW
shape and the concentration is defined by both the core radius and the
shape parameter $\alpha$ in the Einasto approximation. We show that
halos progress through three stages of evolution.  They start as rare
density peaks and experience fast and nearly radial infall that brings
mass closer to the center, producing a highly concentrated halo. Here
the halo concentration increases with increasing halo mass and the
concentration is defined by the $\alpha$ parameter with a nearly
constant core radius. Later halos slide into the plateau regime where
the accretion becomes less radial, but frequent mergers still affect
even the central region. At this stage the concentration does not
depend on halo mass.  Once the rate of accretion and merging slows
down, halos move into the domain of declining concentration-mass
relation because new accretion piles up mass close to the virial
radius while the core radius is staying constant. Accurate analytical
fits are provided.
\end{abstract}


\makeatletter{}\section{Introduction}

The new and upcoming galaxy surveys will be able to measure positions
of millions of galaxies. These large projects will offer good
statistics and will therefore bring the opportunity to test
cosmological models with high accuracy.  To keep up with the
increasing precision of galaxy surveys great care has to be taken when
comparing these observations with cosmological simulations. To begin
with, the surveyed volumes have to be of the same size in observations
and simulations. Therefore, on the one hand one needs to simulate large
volumes \citep[as in][]
{Teyssier2009,KimHorizon2009,Crocce2010,Klypin2011,Prada2012,
DeusSimulation2012,AnguloXXL2012,Jubilee2013} and on the other hand
one wants to keep the resolution high enough to retain reliable
information about the properties of the halos of interest.  The
requirement for mass-resolution stems from the need to identify
and resolve halos which can host galaxies of the observed size to be able to
correctly account for galaxy bias and investigate galaxy clustering \citep[e.g.][]{Nuza2012}.

One of the most basic and important characteristics of halos is their
concentration. Without it we cannot make a prediction of the
distribution of dark matter inside halos. Halo concentration -- its
dependence on mass, redshift, and cosmological parameters -- has been
the subject of extensive analysis for a long period of time
\citep{NFW1997,Jing2000,Bullock2001,Neto2007,Gao2008,Maccio2008,
  Klypin2011,Prada2012,Dutton2014}.

 Already \citet{Jing2000} and
\citet{Bullock2001} found that halo concentration $C$ declines with
 mass. A simple model in \citet{Bullock2001} gave an explanation
for this trend. At late stages of accretion the infalling mass stays
preferentially in the outer halo regions and does not affect much the
density in the center. As the result, the core radius does not change
while virial radius increases with time, which results in growth of
concentration with redshift $C\propto (1+z)^{-1}$ and weak decline
with halo mass $C\propto M^{-0.1}$.

The situation is different at early stages of halo growth when fast
accretion and frequent mergers affect all parts of the halo including
the central region. This results in the growth of both the virial
radius and the core, which in turn leads to a constant concentration
\citep{Zhao2003,Zhao2009}. We will refer to this regime as the plateau
of the concentration -- mass relation.

Recently, \citet{Klypin2011} and \citet{Prada2012} discussed yet
another regime for the halo concentrations: a possible increase of the
concentration for the most massive halos. However,  the situation
with the upturn is not clear. \citet{Ludlow2012} argue that the upturn
is an artifact of halos that are out of
equilibrium. \citet{Dutton2014} did not find the upturn when they fit
the Einasto profile to dark matter density profiles. They suggest that
the upturn may be related with a particular algorithm of finding
concentration used by \citet{Prada2012}.

One of our goals for this paper is to clarify the situation with the
three regimes of the halo concentration. Are the upturn halos out of
equilibrium? Is the upturn an artifact or is it real?

Another complication is related with the choice of analytical
approximations for the halo dark matter density profiles. It is known
that the Einasto profile is more accurate than the NFW, but it has an
extra parameter, which somewhat complicates the fitting procedures of
density profiles. It is traditional in the field to provide fitting
parameters for both Einasto and NFW approximations
\citep[e.g.][]{Gao2008,Dutton2014}. This produces some confusion. For
example, \citet{Dutton2014} compare estimates of concentrations using
three different methods: fitting Einasto and NFW profiles and the
method used by \citet{Prada2012} (the ratio of the virial velocity to
the maximum of the circular velocity). They report some disagreements
between the three methods. We will revisit the issue in the paper.  We
will demonstrate that there is no disagreement between the
methods. They simply have different meanings and apply to different
regimes and different quantities.

In Section~2 we introduce our suite of Multidark simulation and
discuss halo identification.  Relaxed and unrelaxed halos are
discussed in Section~3. Methods of measuring halo concentrations are
presented in Section~4. The evolution of halo properties are discussed
in Section~5. Results for halo concentrations are presented in
Section~6.  In Section~7 we discuss the halo upturn. Comparison with
other published results is presented in Section~8. Possible ways of
using our results to predict density profiles and concentrations are introduced in
Section~9. Summary of our results is presented in Section~10.
Appendix A gives parameters of approximations used in our paper.
Examples of evolution of properties of few halos are presented in the Appendix B.

\makeatletter{}\begin{table*}
 \begin{minipage}{16.cm}
\caption{Numerical and cosmological parameters for the simulations.
  The columns give the simulation identifier, 
  the size of the simulated box in $h^{-1}\,{\rm Gpc}$,
  the number of particles, 
  the mass per simulation particle $m_p$ in units $h^{-1}\,M_\odot$,
  the Plummer equivalent gravitational softening length $\epsilon$  in units of
physical  $h^{-1}\,{\rm kpc}$,
   the adopted values for
  $\Omega_{\rm{Matter}},\, \Omega_{\rm{Baryon}},\, \Omega_{\Lambda}$
  , the clustering at $8 h^{-1}\,{\rm Mpc}$, $ \;\sigma_8$, the spectral
index $n_s$ and the Hubble constant $H_0$ in ${\rm km/s/Mpc}$.}
\begin{tabular}{ l | c | c |  c | c | r | r | r | r | r | r | l | l |l }
\hline  
Simulation & box &  particles  & $m_p$ & $\epsilon$ &
$\Omega_M$ & $\Omega_B$ & $\Omega_{\Lambda}$ & $\sigma_8$ & $n_s$ & $H_0$ & Code & Ref.

\tabularnewline
  \hline   
BigMD27    & $2.5$    &  $3840^3$   &  $2.1 \times 10^{10}$     & $10.0$ &
$0.270$  & $0.047$   &  $0.730$  &  $0.820$  &  $0.95$  &  $70.0$ & {\small GADGET-2} &	1 
\tabularnewline

BigMD29    & $2.5$    &  $3840^3$   &  $2.2 \times 10^{10}$    & $10.0$  &
$0.289$ & $0.047$   &  $0.711$  &  $0.820$  &  $0.95$  &  $70.0$ & {\small GADGET-2} & 1
\tabularnewline

BigMD31    & $2.5$    &  $3840^3$   &  $2.4 \times 10^{10}$    & $10.0$  &
$0.309$  & $0.047$   &  $0.691$  &  $0.820$  &  $0.95$  &  $70.0$ & {\small GADGET-2} &	1
\tabularnewline

BigMDPL  & $2.5$    &  $3840^3$   &  $2.4 \times 10^{10}$  & $10.0$ & 
$0.307$  & $0.048$   &  $0.693$  &  $0.829$  &  $0.96$  &  $67.8$ & {\small GADGET-2} &	1
\tabularnewline

BigMDPLnw  & $2.5$    &  $3840^3$   &  $2.4 \times 10^{10}$  & $10.0$ & 
$0.307$  & $0.048$   &  $0.693$  &  $0.829$  &  $0.96$  &  $67.8$ & {\small GADGET-2} &	1
\tabularnewline

HMDPL  & $4.0$    &  $4096^3$   &  $7.9 \times 10^{10}$  & $25.0$ & 
$0.307$  & $0.048$   &  $0.693$  &  $0.829$  &  $0.96$  &  $67.8$ & {\small GADGET-2} &	1
\tabularnewline

HMDPLnw  & $4.0$    &  $4096^3$   &  $7.9 \times 10^{10}$  & $25.0$ & 
$0.307$  & $0.048$   &  $0.693$  &  $0.829$  &  $0.96$  &  $67.8$ & {\small GADGET-2} &	1
\tabularnewline

  \hline
MDPL    & $1.0$   &  $3840^3$   &  $1.5 \times 10^{9}$  &  $5$ &
$0.307$  & $0.048$   &  $0.693$  &  $0.829$  &  $0.96$  &  $67.8$ & {\small GADGET-2} &	1
\tabularnewline

MultiDark   & $1.0$   &  $2048^3$   &  $8.7 \times 10^{9}$  &  $7.0$ &
$0.270$  & $0.047$   &  $0.730$  &  $0.820$  &  $0.95$  &  $70.0$ & {\small ART} & 2
\tabularnewline

SMDPL   & $0.4$   &  $3840^3$   &  $9.6 \times 10^{7}$  &  $1.5$ &
$0.307$  & $0.048$   &  $0.693$  &  $0.829$  &  $0.96$  &  $67.8$ & {\small GADGET-2} &	1
\tabularnewline

BolshoiP   & $0.25$   &  $2048^3$   &  $1.5 \times 10^{8}$  &  $1.0$ &
$0.307$  & $0.048$   &  $0.693$  &  $0.823$  &  $0.96$  &  $67.8$ & {\small ART} & 1
\tabularnewline

Bolshoi   & $0.25$   &  $2048^3$   &  $1.3 \times 10^{8}$  &  $1.0$ &
$0.270$  & $0.047$   &  $0.730$  &  $0.820$  &  $0.95$  &  $70.0$ & {\small ART} & 3
\tabularnewline
\hline
\multicolumn{8}{l}{1- This paper, 2- \citet{Prada2012}, 3- \citet{Klypin2011}}
\end{tabular}
\label{simtable}

\end{minipage}

\end{table*}

\section{Simulations and Halo identification}
\label{sec:simulations}

Table~1 provides the parameters of our suite of simulations.  Most of
the simulations have $3840^3$ particles. With simulation box sizes
ranging from $250\Mpch$ to $2500\Mpch$ we have the mass resolutions of
$\sim 10^{8} - 10^{10} h^{-1}M_{\odot}$.  Even the moderate resolution
in the $2500 \Mpch$ simulations allows us to resolve a large number of
halos and subhalos that potentially host Milky way like galaxies with
around $\gtrapprox250$ particles. While this resolution does not
provide reliable information about the innermost kpc of a halo, it is
sufficient to estimate key properties such as halo mass, virial radius
and maximum circular velocity and density profiles.

Since the uncertainties of the most likely cosmological parameters in
the newest CMB Planck measurements are so small, it requires special
care to be able to distinguish even small differences in the matter
and halo distribution.  Scaling simulations to other cosmologies, as
suggested in \citet{AnguloWhite2010}, is an important tool but lacks
the required precision with respect to halo properties.  To minimize
the influence of cosmic variance (although it is small in the
simulated volumes) we choose identical Gaussian fluctuations for some
of our simulations. This eliminates the effect of cosmic variance when
comparing the relative differences of the simulations. Initial phases
where different for the MultiDark and Bolshoi simulations done with
the ART code.  When comparing with observations, cosmic variance needs
to be considered despite the size of the surveyed volumes. This is
especially necessary when studying large scales or rare objects.  The
initial conditions based on these fluctuations with identical initial
phases for the simulation are generated with Zeldovich approximation
at redshift $z_{init}=100$.  For simulations BigMDPLnw and HMDPLnw we used
initial power spectra without baryonic oscillations (``no wiggles'').

The simulations we study here have been carried out with {\small
  L-GADGET-2} code, a version of the publicly available cosmological
code {\small GADGET-2} \citep[last described in][]{gadget2} whose
performance has been optimized for simulating large numbers of
particles. We also use the Adaptive Refinement Tree (ART) code
\citep{ART1997,ART2008} for the MultiDark and Bolshoi simulations.
This suite of simulations was analyzed with halo finding codes BDM,
RockStar and FOF. Halo catalogs are provided in the public MultiDark
database \footnote{http://www.multidark.org/}.

In this paper we use results of the spherical overdensity Bound
Density Maxima (BDM) halofinder \citep[see][]{BDM1997,Riebe2013}. The
BDM halo finder was extensively tested and compared with other
halofinders \citep{Knebe2011,ROCKSTAR}. Among other parameters,
BDM provides virial masses and radii.  The virial mass is defined as
mass inside the radius encompassing a given density contrast $\Delta$. We
use two  definitions: the
overdensity $\Delta=200$ relative to the critical density of the
Universe $\rho_{\rm cr}$, and the so called ``virial'' overdensity
$\Delta_{\rm vir}(z)$ relative to the matter density $\rho_{\rm m}$,
which is computed using the approximation of \cite{BryanNorman1998}:

\begin{eqnarray}
  M_{200} &=& \frac{4\pi}{3} \,200  \rho_{\mathrm{cr}} R_{200}^3, \\ 
  \Mvir &=& \frac{4\pi}{3} \,\Delta _{\rm vir}(z) \rho_{\mathrm{m}} \Rvir^3  
  \label{eq:SO_overdensity}
\end{eqnarray}

The BDM halofinder provides a number of properties for each halo. In
addition to $M_{200}$ and $\Mvir$ it lists the maximum value of the
circular velocity $\Vmax:$
\begin{equation}
  \Vmax =\sqrt{\frac{GM(<r)}{r}}\mid_{_{max}}.
\end{equation}
BDM also gives the halo spin parameter $\lambda$, the offset parameter
$X_{\rm off}=\vert \vec r_{\rm centr}-\vec r_{\rm cm}\vert/\Rvir$ -
the distance between halo position (largest potential)
 and the halo center of mass normalized to the virial radius,
and the virial ratio $2K/W-1$, where $K$ is the total kinetic energy
of internal velocities and $W$ is the absolute value of the potential
energy.

To increase the mass coverage of halo properties we combine the
results of the simulations with varying box size and mass
resolution. The smaller box simulations  offer better numerical
resolution, and hence, have a more complete set of objects with low
masses. Rare massive objects are sampled with small statistics because the
simulated volume is smaller.

The halo density profiles and concentrations $C$ depend on redshift
$z$ and halo mass $M$ in a complicated way. Instead of mass and
redshift it is often more physically motivated to relate halo
properties with the height $\nu$ of the density peak in the linear
density fluctuation field smoothed using the top-hat filter with mass
$M$ \citep[e.g.,][]{Prada2012,Diemer2014}. According to a simple
top-hat model of collapse, peaks that exceed density contrast
$\delta_{\rm cr}\approx 1.68$ collapse and produce halos with mass
larger than $M$. In reality, formation of halos is neither spherical
nor simple, but the notion of the peaks in the linear density
perturbation field remains very useful.  It is defined as

\begin{equation}
  \nu \equiv \frac{\deltacr}{\sigma(M,z)},
\end{equation}
where $\sigma(M,z)$ is the rms fluctuation of the smoothed density field:  

\begin{equation}
 \sigma^2(M,z) = 4\pi\int_0^\infty P(k) W^2(k,R_F) k^2 \mathrm{d}k,
\label{eq:sigma}
\end{equation}
where $P(k)$ is the linear power spectrum of perturbations and $W(k,R_f)$ is
the Fourier spectrum of the top-hat filter with radius $R_f$
corresponding to mass $M$.

We start the analysis of our simulations by  presenting a few basic
statistics: the power spectra and mass functions. We have two goals
with these results: (1) show the power of very large and accurate
simulations and (2) demonstrate the smooth transition of halo
properties and the lack of jumps from one simulation to another.
\makeatletter{}\begin{figure}
\includegraphics[width=.51\textwidth]
{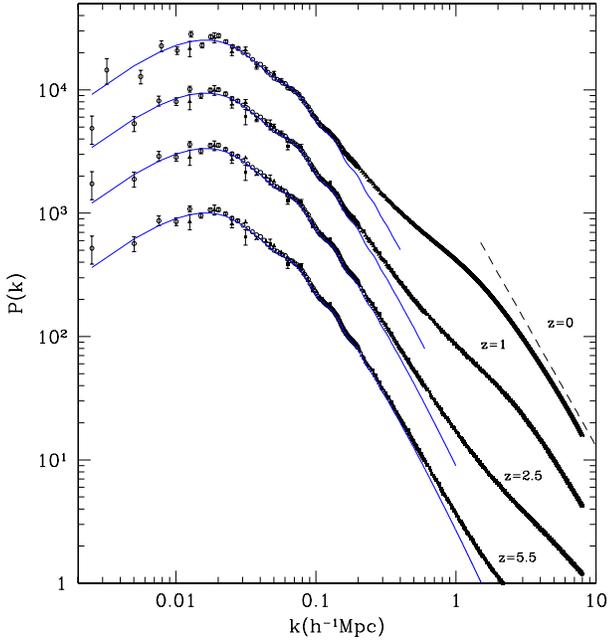}
\vspace{-7.5mm}
\caption{Evolution of the dark matter power spectrum in simulations
  with the Planck cosmological parameters for
  redshifts indicated in the plot. The error bars show errors assuming
  poissonian noise due to the finite number of independent harmonics
  in each bin.  The full (blue) curves show the linear power
  spectra.   Different symbols show BigMDPL
(circles), MDPL (triangles), and SMDPL (squares) simulations.
 The plot shows that there are three regimes of growth of
  perturbations: (1) Linear growth of perturbations on long waves
  gradually shrinks as indicated by the point where the non-linear
  power spectrum starts to deviate upward from the linear theory
  prediction, (2) Mildly non-linear regime where fluctuations grow
  substantially faster than the linear growth, (3) Strongly non-linear
  regime where fluctuations start to approach relatively slow
  self-similar clustering, and the power spectrum is the power-law with
  slope $\sim -2$ indicated by the dashed line in the plot.  }
\label{fig:pk_evol}
\end{figure}

\makeatletter{}\begin{figure}
\begin{center}
\includegraphics[width=.52\textwidth]
{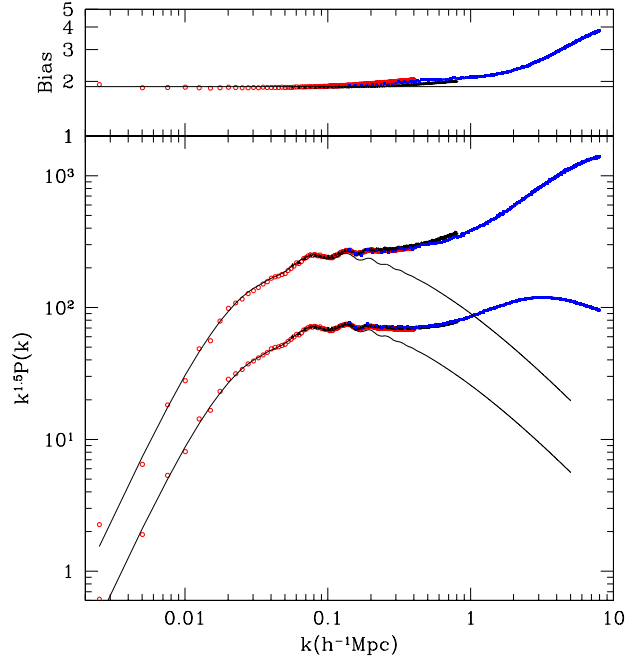}
\vspace{-7.5mm}
\caption{Power spectra (bottom panel) and halo bias (top panel) at
  redshift $z=1$. Different symbols show BigMDPL (circles), MDPL
  (triangles), and SMDPL (squares) simulations.  {\it Bottom panel:}
  Power spectra for dark matter (lower curves) and halos (top curves).
  The power spectra are multiplied by factor $k^{1.5}$ to see more
  clearly the BAO peaks. The full curves show the linear power
  spectrum and the linear power spectrum scaled up with bias factor
  $b=1.95$. Top set of symbols are for dark matter halos with circular
  velocities $\vmax>250 {\rm km/s}$.  {\it Top panel:} Bias factor
  $b(k) =\sqrt{P_{\rm halos}/P_{\rm nonlin DM}}$.  }
\label{fig:pk}
\end{center}
\end{figure}

Figure~\ref{fig:pk_evol} shows the evolution of the power spectrum of
dark matter fluctuations in BigMDPL (circles), MDPL (triangles) and
BolshoiP (squares) simulations. There is a remarkably good agreement
between the simulations in overlapping regions. The plot also
demonstrates with nice clarity the three dynamical regimes of growth
of fluctuations: linear, quasi-linear with growth rates faster than
linear and stable clustering with rates slower than quasi-linear.
The power spectra and bias for dark matter halos and subhalos at redshift
$z=1$ are shown in Figure~\ref{fig:pk}. By plotting the power spectrum
multiplied by $k^{1.5}$ we can see better the 
Baryonic Acoustic Oscillations (BAO).

\begin{figure} \centering
\includegraphics[width=.45\textwidth]
{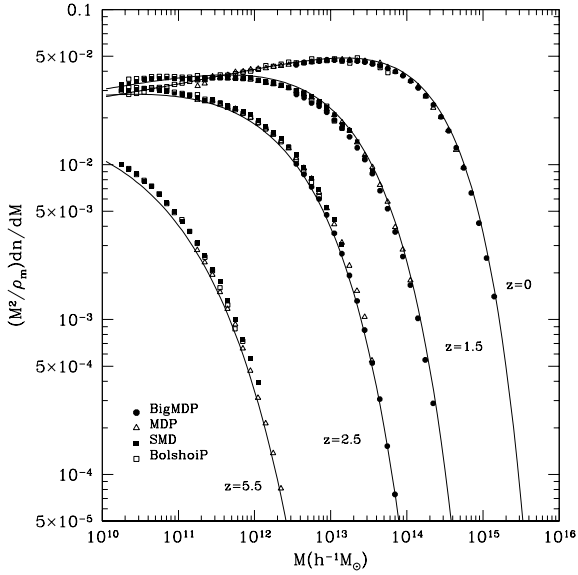}
\caption{
Evolution of the halo mass function in simulations with the Planck
cosmological parameters. Different symbols show results from different
simulations. The full curves correspond to  the \citet{Tinker2008} mass
function. It gives an excellent fit to $z=0$ results and slightly
underestimates the mass function at high redshifts.  }
\label{fig:massF_planck}
\end{figure}

The mass functions of halos at different redshifts in the Planck
cosmology are shown in Figure~\ref{fig:massF_planck}. The
\citet{Tinker2008} mass function provides an excellent fit to $z=0$
results and slightly underestimates the mass function at high
redshifts. The differential mass function 
 shows that the convergence of halo mass
in the simulations is of the order of a few per cent. Compared to the
suite of simulations, the prediction of \citet{Tinker2008} which was
gauged at different cosmologies over-predicts the number of halos by
less than 4\% for halos below $10^{15}\Msunh$. (For definiteness we
employed a quadratic interpolation to get the parameters for $200
\Delta_{crit}$: $A=0.224, \; a=1.67, \; b=1.80, \; c=1.48$ )

\makeatletter{}

\section{Relaxed halos}
\label{sec:Relaxed}
Some dark matter halos may experience significant merging or strong
interactions with environment that distort their density and kinematics, and hence, may bias
the estimation of halo concentrations. To avoid possible biases due to
non-equilibrium effects, we select and separately study halos that
are expected to be close to equilibrium.  Because halos grow in mass
and have satellites moving inside them, there are no truly relaxed
halos. However, by applying different selection conditions, we can
select halos that are less affected by recent mergers and are closer to
equilibrium.

A number of diagnostics have been used to select relaxed halos. These
include the virial parameter $2K/|W|-1$, where $K$ and $W$ are the kinetic
and potential energies, the offset parameter $X_{\rm off}$ (distance
between halo center and the center of mass), and the spin parameter
$\lambda$ \cite[e.g.][]{Neto2007,Maccio2007,Maccio2008,Prada2012}.  

In addition to these three diagnostics,
\citet{Neto2007} and \citet{Ludlow2012,Ludlow2014} also require that the fraction
of mass in subhalos $f_{\rm sub}$ should be small, i.e. $f_{\rm
  sub}<0.1$. This seems to be a reasonable condition, but we decided
not to use it for two reasons.  First, it is redundant: a combination
of cuts in $\lambda$ and $X_{\rm off}$ already remove the vast majority of cases
with $f_{\rm sub}>0.1$. Figure~2 in \citet{Neto2007} clearly shows
this. Second, this condition is very sensitive to resolution. Concentrations are
often measured for halos with few thousand particles. For these halos
one can reliably measure the spin and offset parameters, but detection
of many subhalos is nearly impossible.

\begin{figure}
  \centering
\includegraphics[width=0.48\textwidth]
{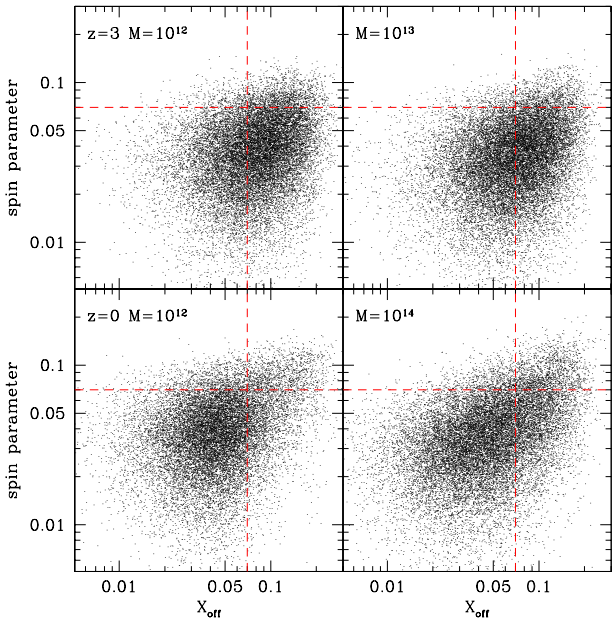}
\caption{Distribution of the offset $X_{\rm off}$ and spin $\lambda$
  parameters for halos in the MDPL simulation at $z=0$ (bottom panels) and
  $z=3$ (top panels), for masses indicated in the
  plot. Dashed lines show the selection of relaxed halos: only halos in the
  low left quadrants are considered to be candidates for relaxed
  halos. In addition to $\lambda$ and $X_{\rm off}$ we also use
  the virial parameter $2K/|W|-1$.}
\label{fig:Xoff}
\end{figure}

Figure~\ref{fig:Xoff} shows the distribution of spin $\lambda$ and
offset $X_{\rm off}$ parameters for halos in the MDPL simulation at
different redshifts and different masses. Note that the distribution
of spin parameters is nearly independent of mass and redshift, a
well known fact. Unlike the spin parameter, the distribution of
$X_{\rm off}$ visibly evolves with time. Dashed lines in the plot show
our condition for relaxed halos:

\begin{equation}
X_{\rm off} < 0.07,\qquad {\rm and} \qquad \lambda < 0.07.
\label{eq:select}
\end{equation}
Depending on redshift and mass, these conditions could select as
``relaxed'' most ($\sim 80$\% for  halos
 $M\lesssim 10^{13}\Msunh$ at $z=0$) or small
($\sim 30$\% for halos $M\gsim 10^{13}\Msunh$ at $z=3$) halo population.

\makeatletter{}

\begin{figure} \centering
\includegraphics[width=0.48\textwidth]{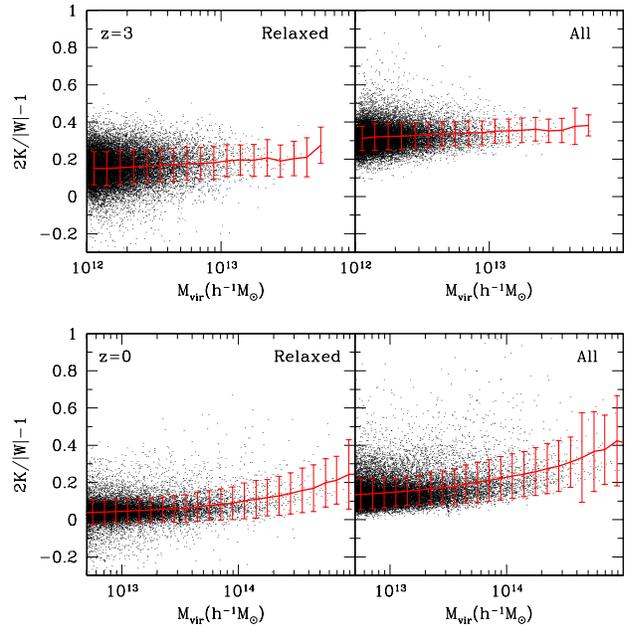}
\caption{Distribution of virial parameters $2K/|W|-1$ for halos in the MDPL
  simulation at $z=0$ (bottom panels) and $z=3$ (top panels).  Points
  show a fraction of all halos to avoid crowding. The full curves show average
  $2K/|W|-1$ values for all halos with the error bars indicating the
  r.m.s. deviations.  Right panels show uncorrected data for all halos. Left
  panels show halos satisfying selection conditions given by
  eq.(\ref{eq:select}), which were corrected for the surface pressure
  effect eq.(\ref{eq:Sp}). Note that the correction brings halos closer to
  equilibrium.}
\label{fig:Virial}
\end{figure}

In addition to the $\lambda$ and $X_{\rm off}$ diagnostics we also use
the virial parameter $2K/|W|-1$. If halos are relaxed and isolated, then
this parameter should be close to zero. Right panels in Figure~\ref{fig:Virial} 
show the  results for halos in the MDPL simulation at $z=0$ and $z=3$.
\makeatletter{}

\makeatletter{}\begin{figure} \centering
\includegraphics[width=0.48\textwidth]{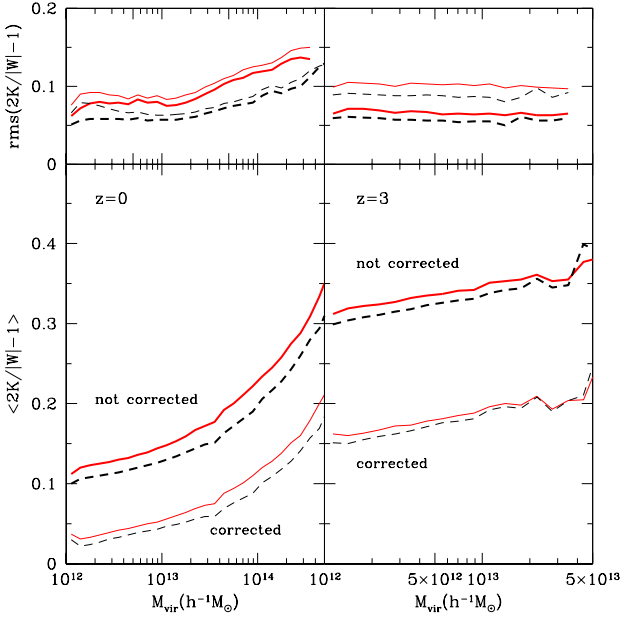}
\caption{ The average virial parameter $2K/|W|-1$ (bottom panels) and
  the r.m.s. deviations from the average (top panels) as the function
  of mass for halos in the MDPL simulation at $z=0$ (left panels) and
  $z=3$ (right panels).  Thick curves are for all halos. Halos
  selected to have small offset parameters and small spin parameters
  as given by eq.(\ref{eq:select}) are shown by dashed curves. Thin
  curves present results corrected for the surface pressure effect
  eq.(\ref{eq:Sp}). The correction brings halos closer to
  equilibrium.}
\label{fig:Virial2}
\end{figure}

As it has been known for a long time that even the average virial
parameter for a population of halos is not zero
\citep[e.g.,][]{Hernquist2001,Shaw2006,Neto2007}. Results show that
the kinetic energy is too large for halos to be in equilibrium.
However, the conclusion that even at $z=0$ most of halos are {\it
  significantly} out of equilibrium is not correct. This is related
with too simplistic application of the virial relation. Because halos
are not isolated, a number of corrections must be taken into account
to assess how far from the equilibrium they really are
\citep{Shaw2006,Knebe2008,Natarajan2011,Power2012}. Here we follow
\citet{Natarajan2011}, who presented the virial relation including
effects of the surface pressure and external force. For a given halo with
mass $\Mvir$, radius $\Rvir$, and kinetic energy $K$, the virial
equation should be written in the following form:

\begin{equation}
2K + W +W_{\rm ext} -S_p = 0,
\end{equation}
where the contribution of the external forces to the potential energy
$W_{\rm ext}$ and the surface pressure term $S_p$ are:
\begin{eqnarray}
W_{\rm ext} &=& -\int_V\rho\vec x \frac{\partial\Phi_{\rm ext}}{\partial x}\mathrm{d}V, \\
S_p        &=& \oint\rho\vec x\cdot\vec v\vec v\cdot \mathrm{d}\vec S =\\
           &=& 4\pi \Rvir^3\rho_{\rm vir}v_r^2 \label{eq:Sp}.
\end{eqnarray}
Here the surface integral in the $S_p$ term is taken over the surface
of a sphere of virial radius $\Rvir$, and $\rho_{\rm vir}$ and
$v_r^2$ are the density and radial velocity dispersion at
$\Rvir$.

We did not try to estimate the correction due to external forces,
though simple estimates indicate that for low concentration halos it
should be similar to $S_p$. In order to estimate $S_p$ we use the halo profiles
 and find both the density and the radial velocity dispersion at
the virial radius. We then estimate the $S_p$ term using
eq.~(\ref{eq:Sp}) and correct the virial parameter $2K/|W|-1$ accordingly.

Left panels in Figure~\ref{fig:Virial} and more detailed
Figure~\ref{fig:Virial2} show effects of surface pressure correction
as well as other corrections on the virial ratio $2K/|W|$.  Typical
value of the surface pressure correction is $\sim 0.1-0.2$, which is
similar to those found by \citet{Shaw2006,Knebe2008} and
\citet{Natarajan2011}. However, the correction depends on mass and on
redshift.  For the most massive halos at $z=0$ the correction is
$\Delta(2K/|W|)\sim 0.15$. It increases to $\sim 0.2$ for the most
massive halos at $z=3$.

 Corrected values of the virial parameter $2K/|W|-1$  are
significantly closer to zero. For example, halos at $z=0$ with mass
$\Mvir< 10^{14}\Msunh$ are very close to equilibrium, which is a
dramatic improvement considering that all of them were considered to
be out of equilibrium without the surface pressure correction. Just as
\citet{Natarajan2011}, we also find that the surface pressure
correction, though an improvement, is not sufficient to bring the most
massive and high-$z$ halos to equilibrium. Other corrections such as
external forces and non-spherical effects are expected to bring those
halos even closer to equilibrium. Unfortunately, it is difficult to estimate
those.

Due to the uncertainty of the estimates of the virial parameter
$2K/|W|-1$, one can apply somewhat loose conditions to select relaxed
halos. For example, following \citet{Neto2007} and \citet{Ludlow2014},
we may have used $2K/|W|< 1.35$. The results in the right panels of
Figure~\ref{fig:Virial} show that this could have been  a reasonable
choice for $z=0$ halos, but not for high redshifts. For example, at
$z=3$, the vast majority of $\Mvir> 10^{13}\Msunh$ halos would be considered
``unrelaxed'' in spite of the fact that they are relaxed (i.e., pass
the $2K/|W|< 1.35$ condition) once the surface pressure correction is
applied.

The combination of a strict criterion for the virial parameter and the lack
of correction for the surface pressure produces an unwanted side
effect in results of \citet{Ludlow2012,Ludlow2014}. Instead of
selecting more relaxed halos, which was their intention, it results in
selection of a biased sample of halos with unusually low
concentrations and infall velocities. 

Instead of applying the surface pressure correction and considering
the fact that most of severely out-of-equilibrium halos are rejected
by the offset and spin parameters, we adopt the  $2K/|W|< 1.5$ selection
applied to the uncorrected virial parameters.

\section{Defining and measuring halo concentrations and  density profiles}
\label{Sec:Define}

\makeatletter{}\begin{figure} \centering
\includegraphics[width=0.48\textwidth]
{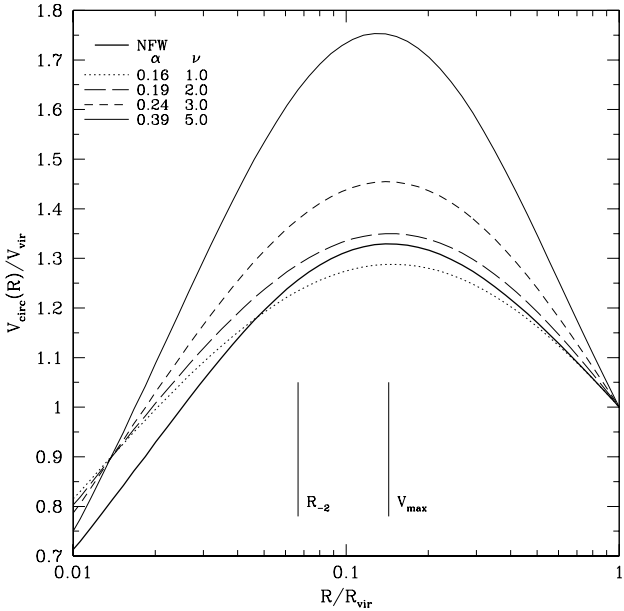}
\caption{Comparison the NFW halo profile with the Einasto profiles
  with different parameters $\alpha$. Halos where fixed to have the
  same virial mass, and the same radius $r_{-2}$ where the slope of
  the density profile is equal to $d\log(\rho)/d\log(R) = -2$.  In
  cosmological simulations the parameter $\alpha$ depends of the peak
  height $\nu$ with larger $\nu$ (and, thus, larger mass $M$) corresponding
  to larger parameters $\alpha$. The ratio of the maximum
  circular velocity to the virial velocity $V_{\rm circ}/V_{\rm vir}$
  is related with halo concentration for any profile.}
\label{fig:Einasto}
\end{figure}

Dark matter halo density profiles are often approximated by the NFW profile \citep{NFW1997}:
\begin{equation}
 \rho_{\rm NFW}(r) = \frac{\rho_0}{x(1+x)^2}, \quad x\equiv \frac{r}{r_{-2}}.
\label{eq:NFW}
\end{equation}
However, halo profiles can substantially deviate
from the NFW shape and are much better approximated with an Einasto profile
\citep{Einasto1965,Navarro2004,Gao2008,Dutton2014}:
\begin{equation}
 \rho_{\rm Ein}(r) = \rho_0\exp\left(-\frac{2}{\alpha}\left[x^\alpha-1 \right]\right),
                 \quad x\equiv \frac{r}{r_{-2}}.
\label{eq:Einasto}
\end{equation}
Here, the radius $r_{-2}$ is the characteristic radius of the halo
where the logarithmic slope of the density profile
$d\log(\rho)/d\log(R)$ is equal to -2. It is often perceived
that the Einasto profile provides accurate fits to the data because it has more free
parameters: three instead of two for NFW. This conclusion is not
correct. Even if the third parameter is fixed (defined by halo mass),
the Einasto profile provides a better fit to simulation results
\citep{Gao2008}. Nevertheless, there is a reason why the NFW profile was used for so
long and why it is still a useful approximation.

\makeatletter{}\begin{figure}
\centering
\includegraphics[width=0.48\textwidth]
{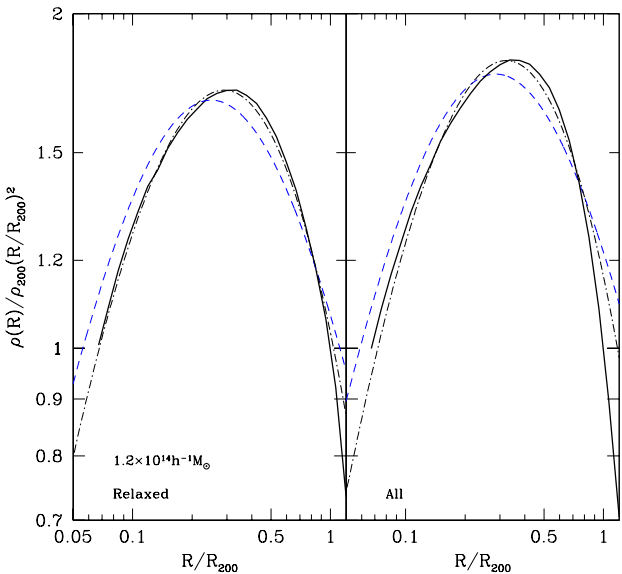}
\caption{Density profiles of halos with mass $M_{200}\approx 1.2\times
  10^{14}\Msunh$ at $z=1.5$ (full curves). Left (right) panels show relaxed (all)
  halos. Dot-dashed curves show Einasto fits, which have the same
  virial mass as halos  in the simulation. The NFW profiles (dashed curves) do not
  provide good fits to the profiles and significantly depend on what part of
  the density profile is chosen for fits. }
\label{fig:ProfileAll}
\end{figure}
\makeatletter{}\begin{figure}
\centering
\includegraphics[width=0.48\textwidth]
{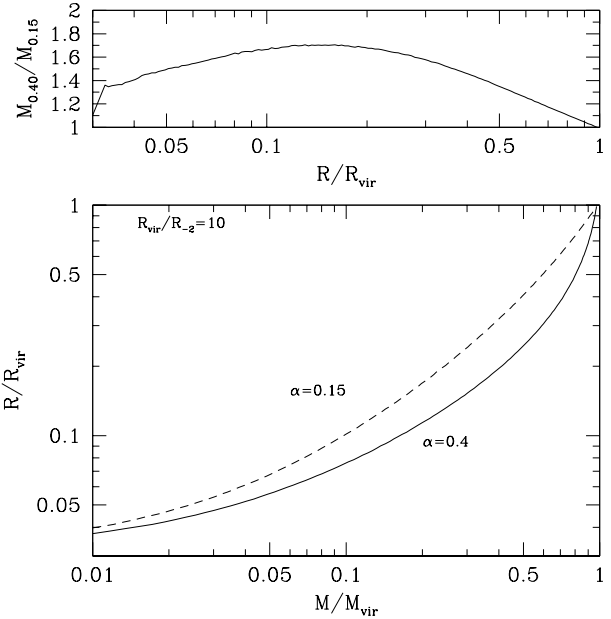}
\caption{Comparison of two Einasto profiles with the same mass
  $\Mvir$, virial radius $R_{\rm vir}$ and core radius
  $r_{-2}=0.1R_{\rm vir}$. The bottom plot shows the radius
  encompassing the same fraction of mass. The top panel presents the
  ratio of masses within the same radius (top panel) for two Einasto
  profiles with shape parameters $\alpha= 0.15$ and $\alpha=
  0.40$. All the statistics indicate that the $\alpha=0.4$ profile is
  more concentrated in spite of having the same formal
  concentration. The plot illustrates that the concentration of the Einasto
  profile depends on both the core radius $r_{-2}$ and the shape parameter
  $\alpha$.}
\label{fig:EinCon}
\end{figure}

The deviations of the NFW profile from $N$-body results are small for
halos that are not massive. For these halos the NFW and Einasto
profiles are very close for a large range of radii $r =(0.01-1)\Rvir$.
This point is illustrated in Figure~\ref{fig:Einasto} that gives
examples of the circular velocity profiles $V^2_{\rm circ}=GM(<r)/r$
for NFW and Einasto profiles. It shows the profiles for halos with the
same virial mass and with the same characteristic radius $r_{-2}$,
which was in this case chosen to be $r_{-2}=\Rvir/15$. To relate
$\alpha$ to $\nu$ we use eq.~(\ref{eq:alphaDutton}). The Einasto
profiles with small $\alpha\approx 0.15-0.18$ parameters that are
typical for halos with peak height $\nu<2$, provide reasonable (though
not perfect) approximations to the NFW with $\lesssim 10\%$ deviations
for radii in the range $R/\Rvir =0.05-1$
\citep[e.g.,][]{Ludlow2012}. The situation is different for much more
massive halos, which are larger peaks of the density field. For
$\nu>2$, the deviations are substantial even for radii $R/\Rvir
=0.1-0.4$ where the circular velocity corresponding to the NFW profile
is $\sim 30$\% below the Einasto profile \citep[see also][]{Gao2008}.

The accuracy of the NFW approximation depends
on  parameters such as halo mass and redshift
\citep[e.g.,][]{Navarro2004}. \citet{Gao2008} and \citet{Dutton2014}
argue that the second parameter $\alpha$ in the Einasto profile
depends only on the amplitude of perturbations $\sigma(M)$, where $M$
is the halo mass.  \citet{Gao2008} provide the following approximation
for the dependence of $\alpha$ on the amplitude of perturbations:
\begin{equation}
  \alpha = 0.155 +0.0095\nu^2.
\label{eq:alphaDutton}
\end{equation}

The observed large deviations from the NFW profile for high-$\nu$
peaks pose a number of problems for the estimates of halo
concentrations because the concentrations are routinely estimated
using NFW fits
\citep[e.g.,][]{Neto2007,Duffy2008,Diemer2014,Dutton2014}.
Figure~\ref{fig:ProfileAll} shows an example that highlights the
problem. At $z=1.5$ we select $\sim 3000$ halos with mass $M_{200}=
(1-1.5)\times 10^{14}\Msunh$ in the BigMDPL simulation. These halos
correspond to high density peaks of $\nu\simeq 3.6$. We then construct
the median halo density profiles for these
halos and present them in  Figure~\ref{fig:ProfileAll} for all and
for relaxed halos. The plots demonstrate that the Einasto
approximation provides remarkably accurate fits for the measured
profiles with deviations less than 1\% for the relaxed halos. The NFW
profile is much less accurate. However, the real problem with the NFW
fits is the systematic errors, not the random errors. The NFW fits
systematically overpredict the halo concentration by (10--20)\% for
the high-$\nu$ halo density profiles shown in
Figure~\ref{fig:ProfileAll}. This and other systematic effects were
recently discussed by \citet{MeneghettiRasia2013} and
\citet{Dutton2014}.

One may think that using the Einasto profiles to estimate
concentrations would produce much better results.  Unfortunately, the
Einasto profile has its own critical issue when it comes to
concentration. For the NFW profile the ratio $R_{\rm
  vir}/r_{-2}$ uniquely defines the density profile for given halo
mass, and therefore it is a good measure for halo
concentration. Yet, for the Einasto profile such a ratio is not the
 concentration because for the same $R_{\rm vir}/r_{-2}$ ratio,
halos with larger $\alpha$ are clearly denser, and thus are more
concentrated.  In order to understand why is that, we need to step
away from the problem of fitting profiles of cosmological dark matter
halos and discuss what is concentration.

The common-sense notion of concentration is that for two objects of
fixed mass and radius, the object with a more dense center is more
concentrated.  If we consider an astronomical quasi-spherical object
such as a globular cluster, an elliptical galaxy, or a dark matter
halo, then a more concentrated object is the one that has denser
central region and less dense outer halo.
Keeping in mind this general notion of concentration, we now look
again at the NFW and Einasto profiles presented in
Figure~\ref{fig:Einasto}. All the profiles in the plot have the same
virial mass, virial radius and $r_{-2}$. However, they all have different
concentrations because they have different masses inside the central
$\sim 0.1\Rvir$ radius.  This happens because  the ``shape
parameter'' $\alpha$ in the Einasto profile affects not only the shape
of the profile, but also its concentration.

\makeatletter{}\begin{figure*} \centering
\includegraphics[width=0.48\textwidth]
{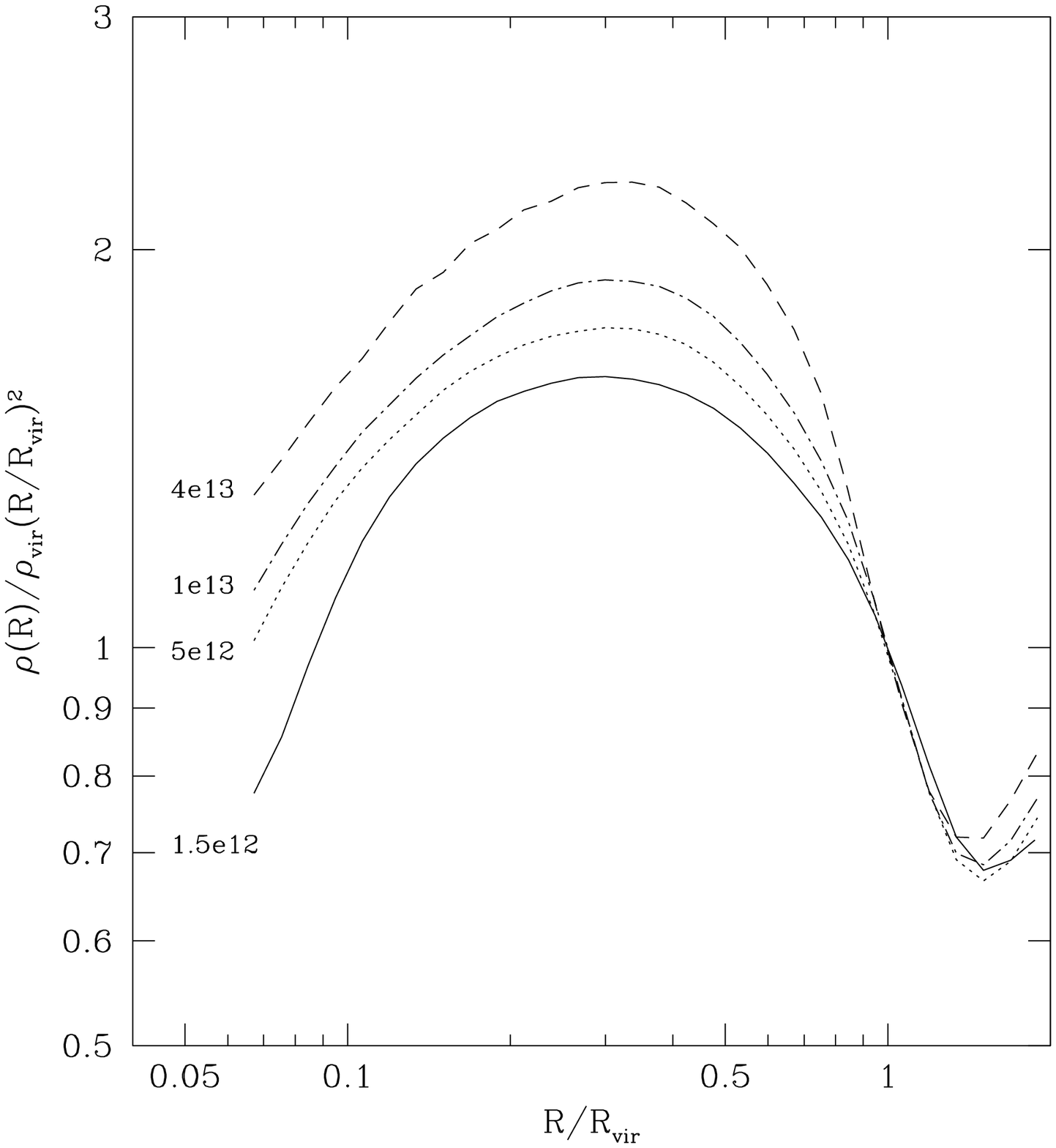}
\includegraphics[width=0.48\textwidth]
{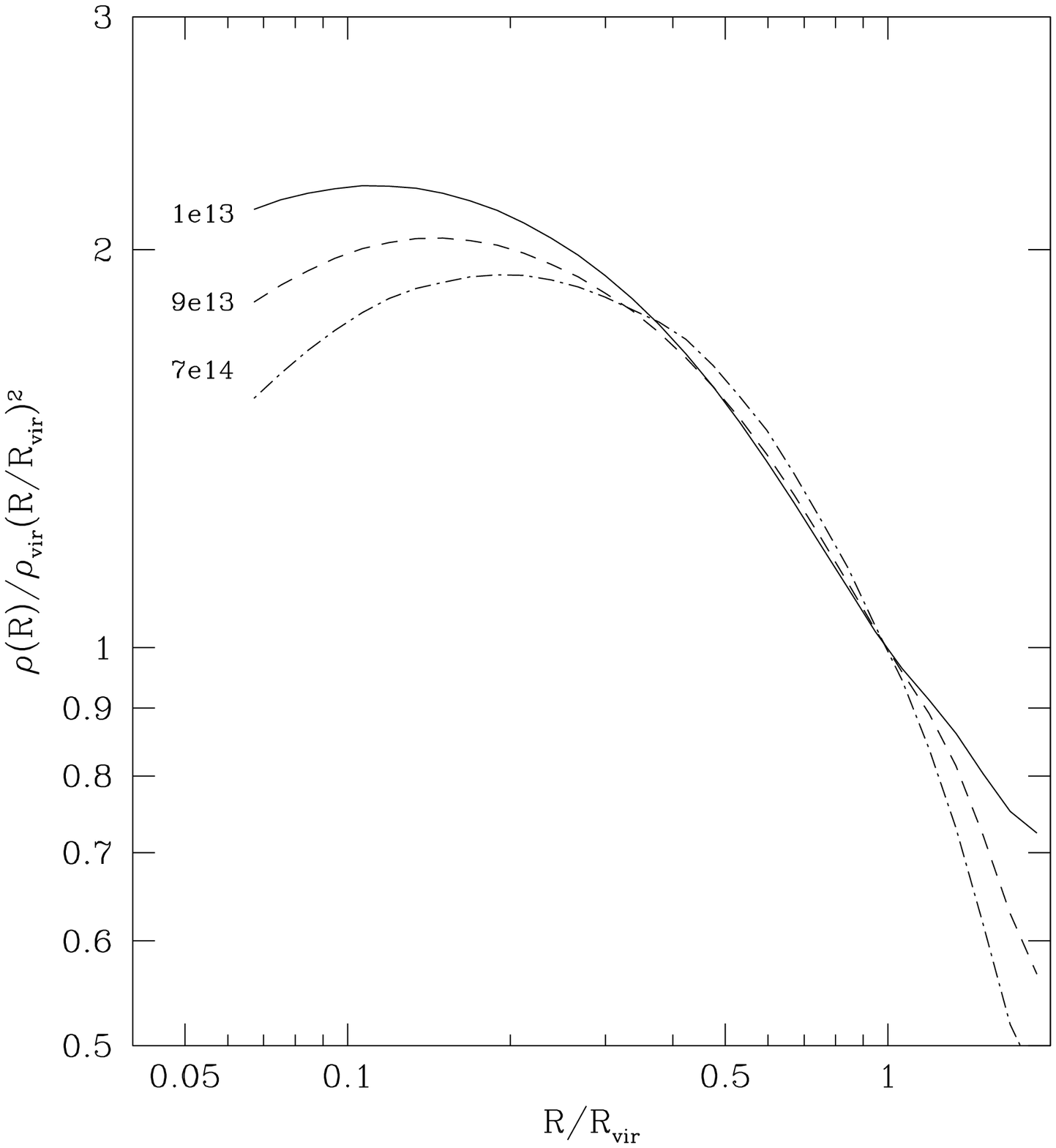}
\caption{Median density profiles of relaxed halos at different
  redshifts and masses in the MDPL simulation. Profiles are
  normalized to have the same density at the virial radius. The left
  panel is for halos at $z=3$: halos with larger mass are clearly more
  concentrated than halos with smaller masses. Similar to Einasto
  profiles in Figure~\ref{fig:Einasto}, value of $r_{-2}$ radius
  almost does not change with  halo mass, which indicates that the
  increase in the concentration is mostly due to the increase in shape
  parameter $\alpha$.  The right panel shows profiles of halos at
  $z=0$. Note that the trend with mass is different: more massive
  halos are less concentrated and $r_{-2}$ radius decreases with
  decreasing mass.}
\label{fig:DensityProf}
\end{figure*}

Figure~\ref{fig:EinCon} illustrates the point. Here we analyze two
Einasto profiles with the same total mass and the same ratio
$\Rvir/r_{-2}=10$. The only difference is the shape parameter: $\alpha
=0.15$ and $\alpha =0.40$. For these density profiles
Figure~\ref{fig:EinCon} shows two conventional measures of
concentration: radius of a given fraction of mass (bottom panel) and
fraction of mass within given radius (top panel).  On all accounts,
the profile with larger $\alpha$ is more concentrated. For example,
the radius of (1/3) of the virial mass is about 30\% smaller for the
$\alpha =0.40$ model. 

To summarize, for the Einasto profile the $\Rvir/r_{-2}$ ratio is not
the concentration. The real concentration depends on both
$\Rvir/r_{-2}$ and $\alpha$.

Following \citet{Prada2012} we use the ratio of the maximum circular
velocity to the virial velocity $V_{\rm max}/V_{\rm vir}$ as a
{\it profile-independent measure of halo concentration}. The larger is the
ratio, the larger is the halo concentration regardless of any particular
analytical approximation of the profile.

It is convenient to write the circular velocity profile $\Vcirc(r)$ for
the NFW and Einasto approximations in the following way:
\begin{eqnarray}
 \left(\frac{\Vcirc(r)}{V_{\rm vir}}\right)^2 &=& \frac{C_E}{x} \frac{f_E(x,\alpha)}{f_E(C_E,\alpha)} 
      = \frac{C}{x} \frac{f(x)}{f(C)},\label{eq:Vone} \\
   f(x) &=& \ln(1+x) -\frac{x}{1+x},\\
   f_E(x,\alpha) &=& e^{\frac{2}{\alpha}}\int_0^xx^2e^{-\frac{2}{\alpha}x^\alpha}dx,\\
    x&\equiv& \frac{r}{r_{-2}}, \quad
    C  = \frac{\Rvir}{r_{-2}}.
\label{eq:Vcirc}
\end{eqnarray}
Here, functions $f(x)$ and $f_E(x)$ define the mass profiles for  NFW
and Einasto approximations correspondingly, and $C$ and $C_E$ are formal halo
concentrations.  Using these relations one can find the radius $x_{\rm
  max}$ at the maximum of the circular velocity, $V_{\rm max}$.

Once $V_{\rm max}/V_{\rm vir}$ ratio is measured in simulations, it
can be used to estimate the formal concentration
$\Rvir/r_{-2}$. For the Einasto approximation one can use one of the two relations in eq.~(\ref{eq:Vone}):
\begin{eqnarray}
  \frac{\Vcirc^2(r)}{V^2_{\rm vir}} &=& \frac{C_E}{x_{\rm max}} 
  \frac{f_E(x_{\rm max},\alpha)}{f_E(C_E,\alpha)},\phantom{mmm} \label{eq:Vtwo} \\
  && \phantom{mmm} x_{\rm  max} \approx 3.15\exp(-0.64\alpha^{1/3})           
\label{eq:Vthree}
\end{eqnarray}
And for the NFW profile:
\begin{equation}
 \frac{\Vcirc^2(r)}{V^2_{\rm vir}}
      = \frac{C}{x_{\rm max}} \frac{f(x_{\rm max})}{f(C)}, \quad x_{\rm max} = 2.163\label{eq:Vfour}
\end{equation}

\makeatletter{}\begin{figure} \centering
\includegraphics[width=0.49\textwidth]
{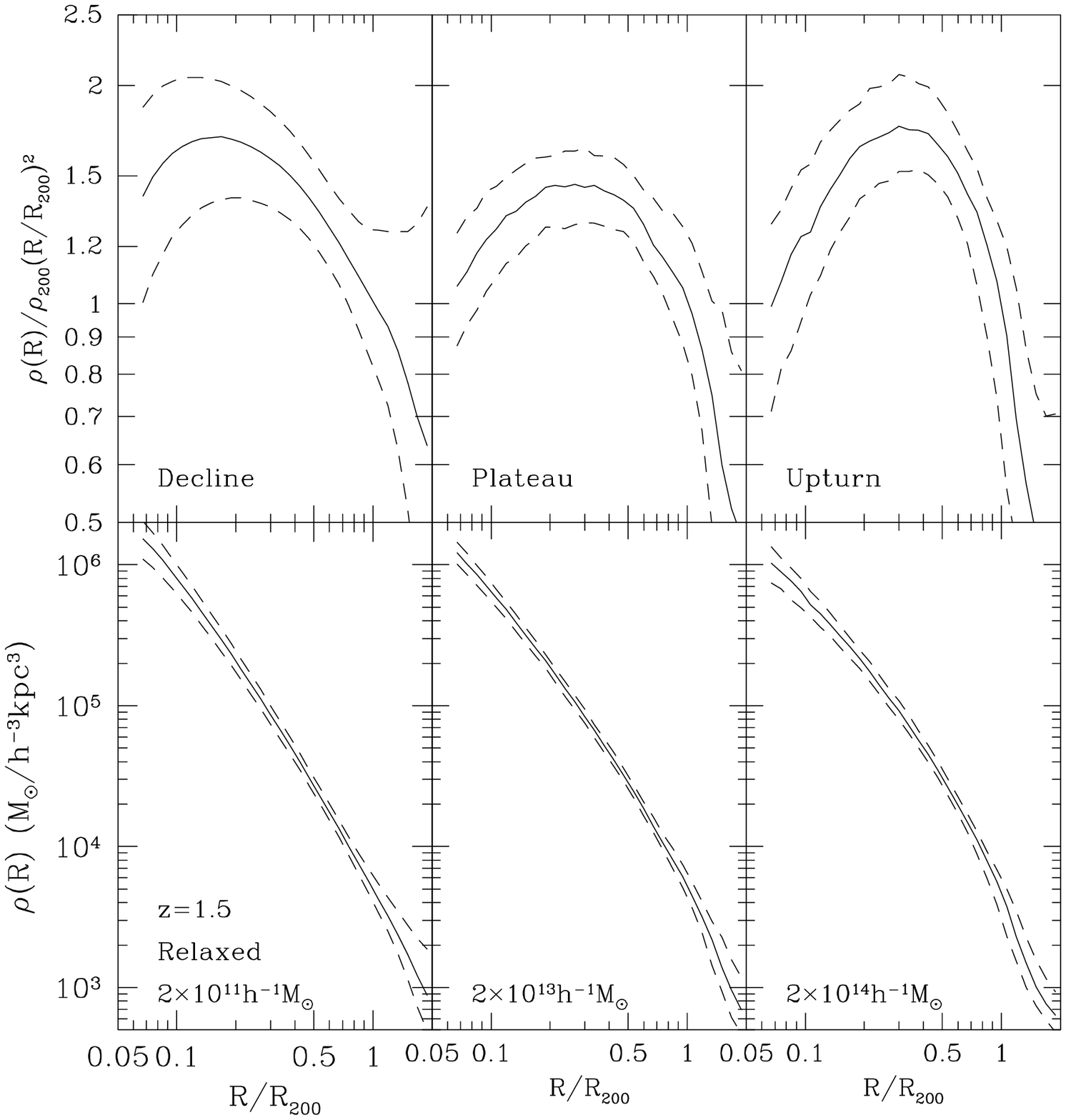}
\caption{Median halo profiles (full curves) of relaxed halos in
  different dynamical regimes. All halos are selected at
  $z=1.5$. Dashed curves show $1\sigma$ deviations from the median.
  Left panels show halos in the declining part for concentration-mass
  $c(M)$ relation. The middle panels are for halo close to the minimum
  of $c(M)$, and the right panels show halos in the upturn. Top panels
  show density profiles multiplied by the square of radius and
  normalized to the density at the virial radius. Halos in the
  declining $c(M)$ regime are more concentrated than halos in the
  ``plateau'', which can be judged by slightly larger densities in the
  central $R\lesssim 0.1R_{200}$ region and by smaller radius where
  $R^2\rho(R)$ reaches the maximum. Halos in the upturn are denser
  than at the plateau, but in a different way: they have larger
  density in intermediate radii $R/R_{200}\approx (0.2-0.5)$ and
  the radius of their peak of $R^2\rho(R)$ does not change.}
\label{fig:ProfileRel}
\end{figure}

It is convenient to cast the $V_{\rm max}/V_{\rm vir}$ ratio into the
concentration using the NFW profile. For this we use
eq.~(\ref{eq:Vfour}) to convert the velocity ratio into concentration
$C$.  For halos that can be approximated by the NFW profile -- and
vast majority of them are -- this gives us the familiar relation
$C=\Rvir/r_{-2}$. For high-$\nu$ halos that are not approximated by
the NFW profile, this still gives a measure of concentration: the
larger $C$ the more  dense is the central region of these
halos. Because the mapping from the $V_{\rm max}/V_{\rm vir}$ ratio to
the concentration $C$ is monotonic, it always can be inverted to recover
$V_{\rm max}/V_{\rm vir}$ and then to find $\Rvir/r_{-2}$ for halo with given
parameter $\alpha$.

The BDM halofinder provides measurements of $\Vmax$ and $V_{\rm vir}$ for
each halo, which are converted to estimates of concentration $C$. When
studying dependence of concentration on mass or $\Vmax$, we bin halos
according to their mass or $\Vmax$. Halos in each bin (typically many
thousands) are ranked by their values of $C$ and median values and
deviations from the mean are found.

Similar procedure is used for dark matter profiles or profiles of the
radial infall velocities and velocity anisotropy parameter $\beta$. We
normalize radii to the virial radius of each halo.  Then profiles are
binned using constant bin size in logarithm of radius $\Delta
R/\Rvir=0.05$. Radii are binned from $\sim 5$\% of $\Rvir$ to
$2\Rvir$. For each radial bin, values from individual halos are ranked
from the smallest to the largest and medians and deviations are found.

\section{Halo profiles at different redshifts}
\label{sec:Devolve}

Some of the halos in our simulations are resolved with millions of
particles, but a vast majority have just few thousands. For this paper
our goal is to study these numerous halos that are resolved from
modest radius $\sim 0.05\Rvir$ to $2\Rvir$. Simulations provide us  the
density profiles and also profiles of velocity dispersions, velocity
anisotropies and radial infall velocities. Because of the wealth of
information, we cannot show all combinations of halo
profiles. Instead, we show only representative results.

One of the interesting issues is the structure of high-$\nu$ halos and
the upturn in halo concentrations. In order to avoid possible
complications with on-going major-mergers, we start the analysis with
density profiles of relaxed halos. We select halos in a narrow range
of masses and normalize distances by the virial radius and normalize
densities by the density at the virial radius. Even with the narrow
mass bins the number of halos in each bin is so large that the
statistical errors of median values are very small. This is the reason
why we do not show any statistical error bars in our plots.

The right panel Figure~\ref{fig:DensityProf} shows density profiles
for relaxed halos at $z=0$.  Even without fitting profiles with either NFW or
Einasto profiles, the figure shows the well known trend with mass:
more massive halos are less concentrated than the less massive
ones. This can be seen by comparing densities at, say $R\approx
0.1\Rvir$. There is another trend: the radius at which the density
declines as $\rho\propto R^{-2}$ (radius where curves in the plot are
horizontal) increases with increasing mass. While the halos in the
plot have large mass, in the sense of peak height they are not large:
$\nu =1-2.8$. 

Halos at larger redshifts demonstrate strikingly different
profiles. The left panel in Figure~\ref{fig:DensityProf} presents
halos at $z=3$.  Here the trend with mass is very different with more
massive halos being more concentrated. There is a number of ways to
demonstrate this. For example, inside a given fraction of virial
radius $R/\Rvir$ larger halos have larger fraction of halo mass. Also
the radius containing a given fraction of mass (another measure of
concentration) is smaller for more massive halos. Another change in
the halo profiles is the radius with the log-log slope -2 for density
profiles. For these halos there is very little dependence of
$\Rvir/R_{-2}$ on halo mass.

\makeatletter{}\begin{figure}
\centering
\includegraphics[width=0.50\textwidth]{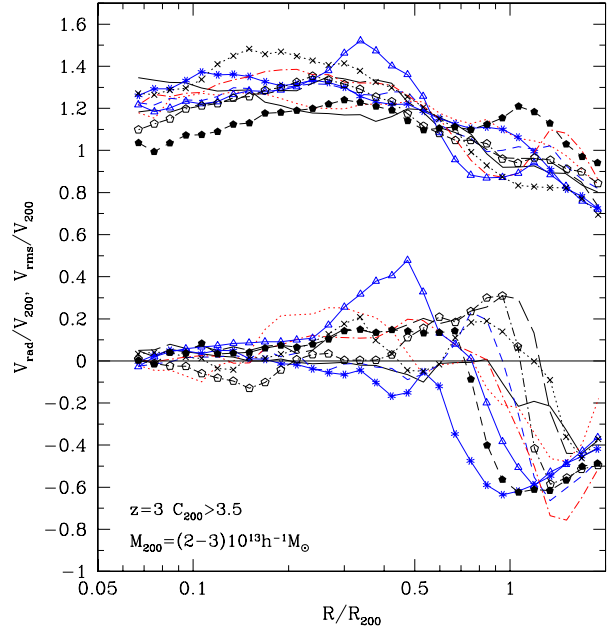}
\caption{
  Examples of radial velocity (bottom curves) and 3D rms velocity (top
  curves) profiles for halos at the upturn of concentration-mass
  relation. Shown are results for 10 halos at $z=3$ with virial mass
  $\M200=(2-3)10^{13}\Msunh$ and concentrations $C_{200}>3.5$. While
  the outer halo regions show substantial infall velocities, halo
  centers $R<0.3R_{200}$ demonstrate clear signs of virialization and
  equilibrium.}
\label{fig:InfallSample}
\end{figure}

Note that the halos at $z=3$ shown in the left panel of
Figure~\ref{fig:DensityProf} represent very large density peaks:
$\nu=2.6-4.5$.  Indeed, it is mostly the height of the peak that
defines the change in the structure of halo profiles. As we will see
later, there is a dependence with the redshift, but this appears to
be a weaker factor shaping the the structure of halos with peak height
being the dominant effect. We can demonstrate this by selecting halos
at the same redshift but with different $\nu$.

\makeatletter{}\begin{figure}
\centering
\includegraphics[width=0.47\textwidth]{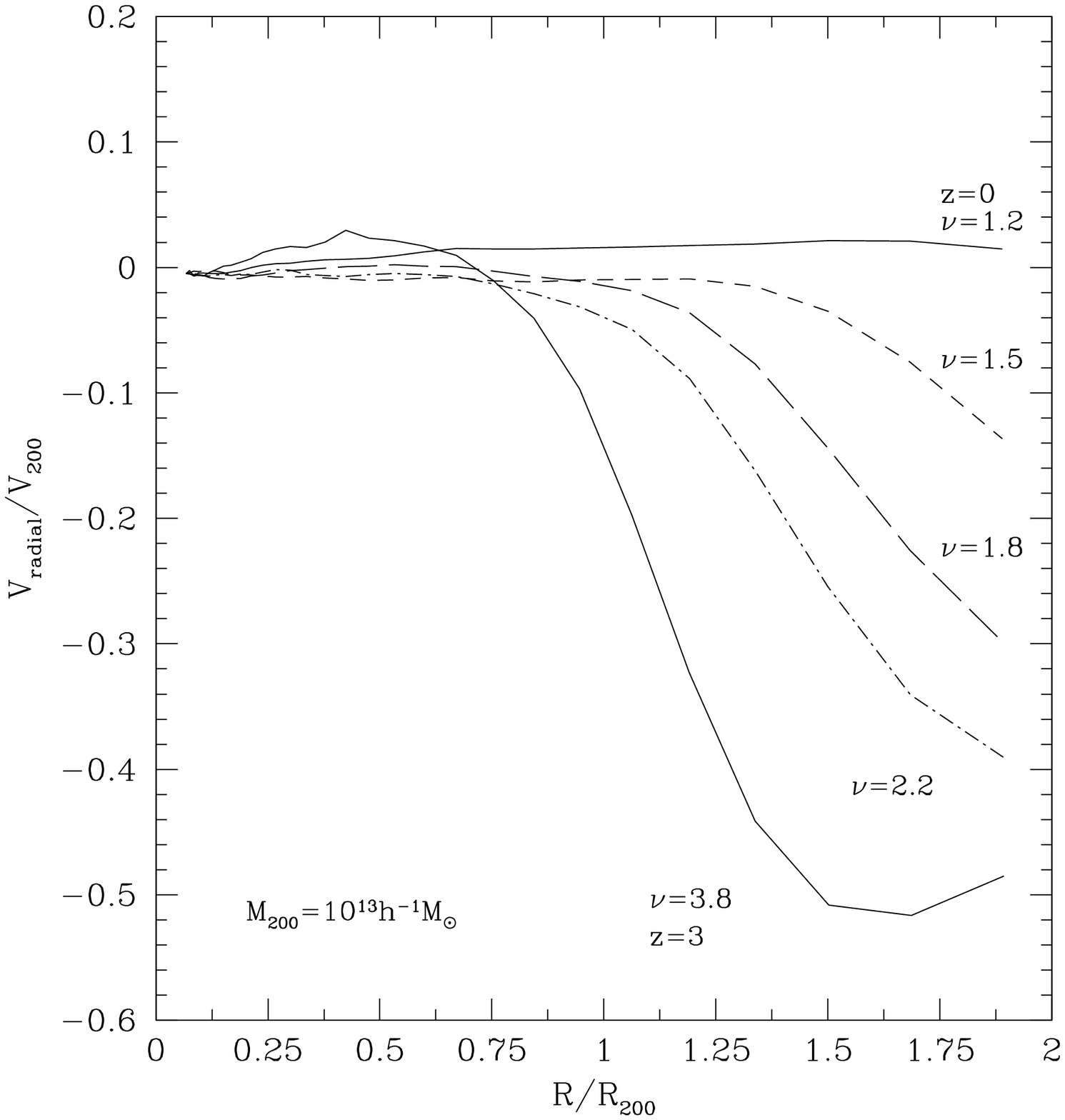}
\caption{Infall velocity for halos with mass
  $M_{200}= 10^{13}\Msunh$ at different redshifts. 
  Median radial velocity profiles of relaxed halos at redshifts
  $z=0,0.5,1,1.5,3$ (from top to bottom). Halos with a constant mass
  represent density peaks of different height $\nu$ as indicated in
  the plot.  There are large infall velocities
  $\langle V_{radial}\rangle \approx 0.5V_{200}$ just outside the
  virial radius for high-$\nu$ halos, which changes to almost no
  infall for $\nu\lesssim 1$. The inner regions of all halos clearly
  indicate relaxation with the median radial velocity being very close
  to zero for both relaxed and unrelaxed halos.}
\label{fig:Infall}
\end{figure}

For that we select relaxed halos at $z=1.5$. The first sample of halos
has mass $\Mvir=2\times 10^{11}\Msunh$ and $\nu=1.1$. These halos are
in the regime of declining concentration-mass relation $C(M)$. Halos
in the regime of the flat part of the $C(M)$ are represented by halos
with $\Mvir=2\times 10^{13}\Msunh$ and $\nu=2.5$ and the upturn of the
$C(M)$ is represented by $\Mvir=2\times 10^{14}\Msunh$ and $\nu=4.1$.
Profiles of these three types of halos are shown in
Figure~\ref{fig:ProfileRel}.  Halos in the plateau of the $C(M)$ have
the lowest concentration and halos in the declining and upturn regimes
are more concentrated. There is also a change in the shape of the
density profiles.  For example, the $r^2\rho(r)$ curves are broader
for plateau halos and are more narrow for the upturn halos. 

\makeatletter{}\begin{figure}
\centering
\includegraphics[width=0.48\textwidth]{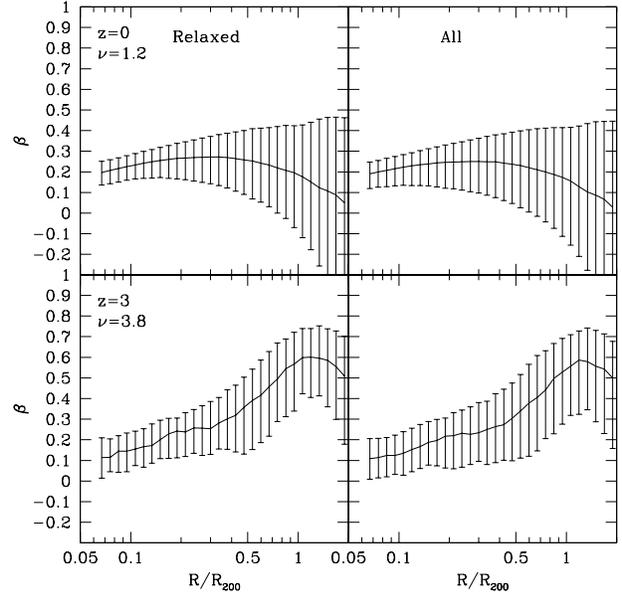}
\caption{Velocity anisotropy
  parameter $\beta$ for halos with virial mass $M_{200}=
  10^{13}\Msunh$ for redshift $z=0$ (top panels) and $z=3$ (bottom
  panels). Relaxed halos were used for plots in left panels and all
  halos were used for right panels. Error bars indicate $1\sigma$
  deviations from the mean.  At $z=0$ these halos represent low-$\nu$
  peaks in density fluctuations. They are in ``normal'' declining
  $c(M)$ regime. Halos have slightly radial orbits of dark matter
  particles with nearly constant $\beta \approx 0.2$. At $z=3$ the
  halos with the same mass are high-$\nu$ peaks in the upturn regime
  of $c(M)$. Orbits are very radial in the outer halo regions with
  $\beta \approx 0.5$ at $R_{200}$. Note that there is almost no
  difference between all and relaxed halos in spite of the fact that
  at $z=3$ only 1/3 of halos are considered relaxed.}
\label{fig:Beta}
\end{figure}

Figure~\ref{fig:ProfileRel} also shows rms deviations from the median
profiles of halos. There is not much difference between different
populations of halos. In this respect rare halos at the upturn are not
more violent and do not have larger spread of densities as compared
with ``normal'' halos of the declining $C(M)$ population.
However, many other halo properties do depend on $\nu$.

 We start our analysis of velocity profiles of halos by presenting
  examples of radial and rms velocites of high peaks halos.
  Figure~\ref{fig:InfallSample} presents velocity profiles (both
  radial and total rms) for all 10 halos in the simulation SMDPL at
  redshift $z=3$ with masses $\M200=(2-3)10^{13}\Msunh$ and
  concentrations $C_{200}>3.5$. The halos are at the upturn. They
  clearly show strong infall velocities in the outer $R>0.5R_{200}$
  regions. However, the central $R\lesssim 0.3R_{200}$ regions, that
  define halo concentrtion, are quiet and have small radial
  velocities. Another indicator of relaxation is a large ratio of
  random velocites to the bulk radial velocities.

 The average radial velocity profiles and the velocity anisotropy
  parameters $\beta$ are shown in Figures~\ref{fig:Infall} and
  ~\ref{fig:Beta} for halos with different peak heights $\nu$. 
Nearly zero average infall velocities $V_{\rm radial}\approx 0$ indicate that
central regions of halos $R\lesssim R_{200}$ are close to
equilibrium at all times.  This is true even for $z=3$ halos that
are at the upturn regime of the $C(M)$ relation. This confirms our
estimates of the virial parameter $2K/|W|-1$ in
Sec.~\ref{sec:Relaxed}: corrected for the surface pressure the virial
parameter is close to zero. 

Negative $V_{\rm radial}$ values at radii larger than the virial
radius indicate that matter falls into the halos resulting in its mass
growth. However, the magnitude of the infalling velocities drastically
depends on $\nu$. It is nearly a half of the circular velocity at the
virial radius for $\nu>3$ and goes to almost zero for $\nu \lesssim 1$
\citep{Prada2006,Cuesta2008,Diemer2013}.
Large infall velocities on high-$\nu$ halos are accompanied by very
radial velocity anisotropy as demonstrated by right panels in
Figure~\ref{fig:Beta}.

In order to characterize the density profiles of halos at different
redshifts, masses and different $\nu$, we stack profiles of halos
binned by mass and normalize radii by virial radius of each halo. The
number of halos in each mass bin is many hundreds and often
thousands. We use only halos with more than 5000 particles. We then
fit the Einasto profile eq.(\ref{eq:Einasto}) using {\it two} free
parameters: $r_{-2}$ and $\alpha$. The third parameter is fixed to
produce the total mass corresponding to the mass of selected halos. We
note that this is not what is usually done
\citep[e.g.,][]{Maccio2007,Gao2008,Dutton2014}.  Usually all three
parameters are considered as free. This cannot be correct once the
total mass is fixed by selecting halos by mass.

\begin{figure*}
\centering
\includegraphics[width=0.48\textwidth] {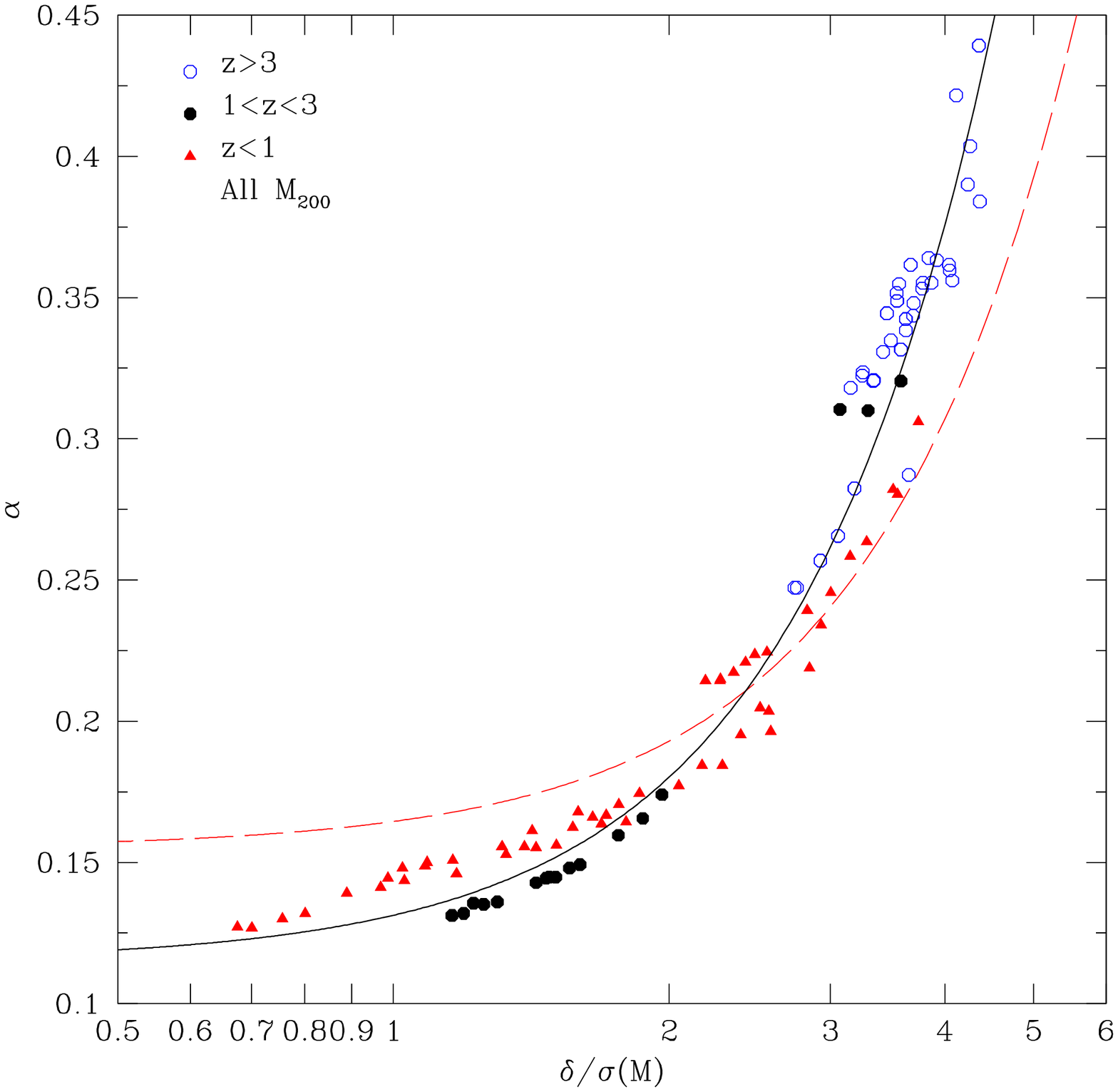}
\includegraphics[width=0.48\textwidth] {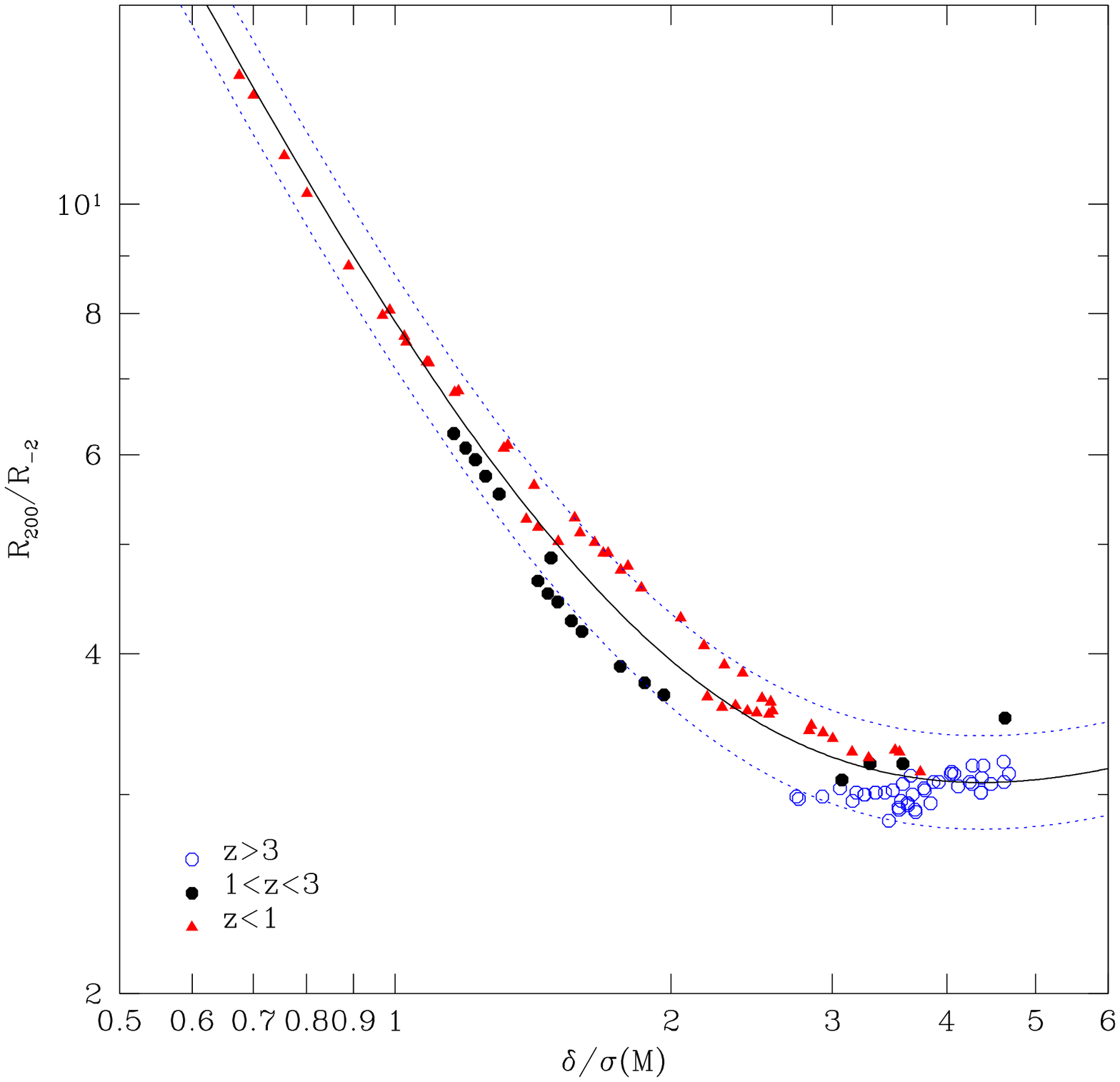}
\caption{Dependence of the structural parameter $\alpha$ (left panel)
  and the ratio of the virial radius $R_{200}$ to the Einasto core
  radius $r_{-2}$ (right panel) on the peak height
  $\nu= \delta_{\rm cr}/\sigma(M,z)$.  All halos in simulations with
  the Planck cosmology were used for this analysis. The full curves
  shows the analytical approximations eqs.(\ref{eq:alpha}-\ref{eq:cc})
  The dotted curves in the right panel indicate 10\% deviations from
  it. The dashed curve on the left panel shows approximation
  eq.(\ref{eq:alphaDutton}) by \citet{Gao2008}.}
\label{fig:AlphaSigma}
\end{figure*}

Parameters $\alpha$ and $r_{-2}$ of all halos in the Planck cosmology
simulations SMDPL, MDPL, and bigMDPL are presented in
Figure~\ref{fig:AlphaSigma}.  Dependence of $\alpha$ and
$R_{200}/r_{-2}$ on the peak height can be approximated with the following
expressions:

\begin{eqnarray}
  &\alpha& = 0.115  +0.0165\nu^2, \label{eq:alpha}\\
  &R_{200}/r_{-2}& = 6.5\nu^{-1.6}\left(1+0.21\nu^2\right),
\label{eq:cc}
\end{eqnarray}
For relaxed halos we find that $\alpha$ parameter is slightly smaller
for high-$\nu$ halos and can be approximated as:
\begin{equation}
  \alpha = 0.115  +0.0140\nu^2.
\label{eq:alphaRelax}
\end{equation}


\makeatletter{}

Evolution of halo density profiles for few halos around $z\approx 3$ are given in Appendix B.
\section{Halo Concentrations: evolution with redshift}
\label{sec:Cevolve}
\begin{figure*}
\centering
\includegraphics[width=0.49\textwidth]
{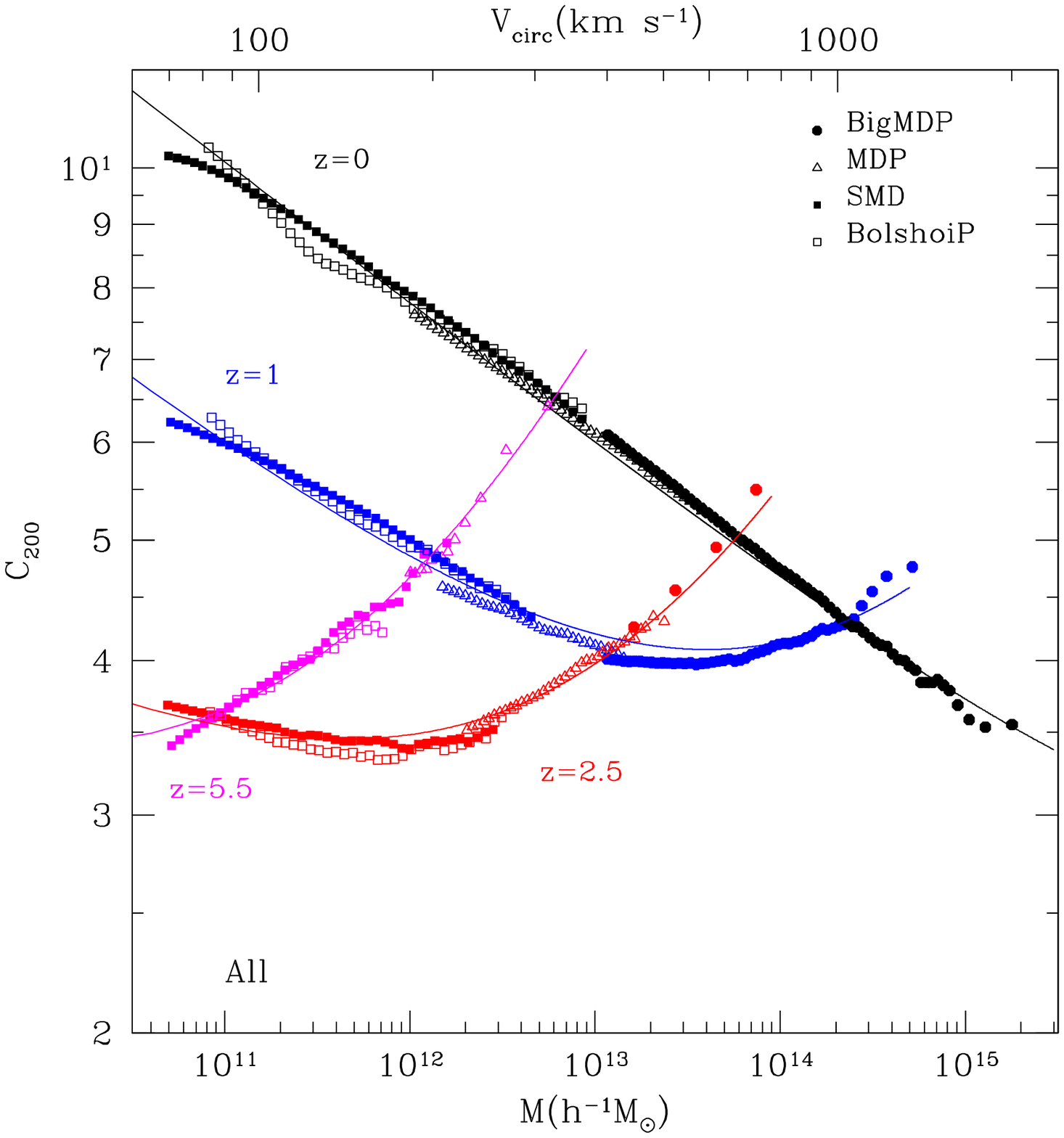}
\includegraphics[width=0.49\textwidth]
{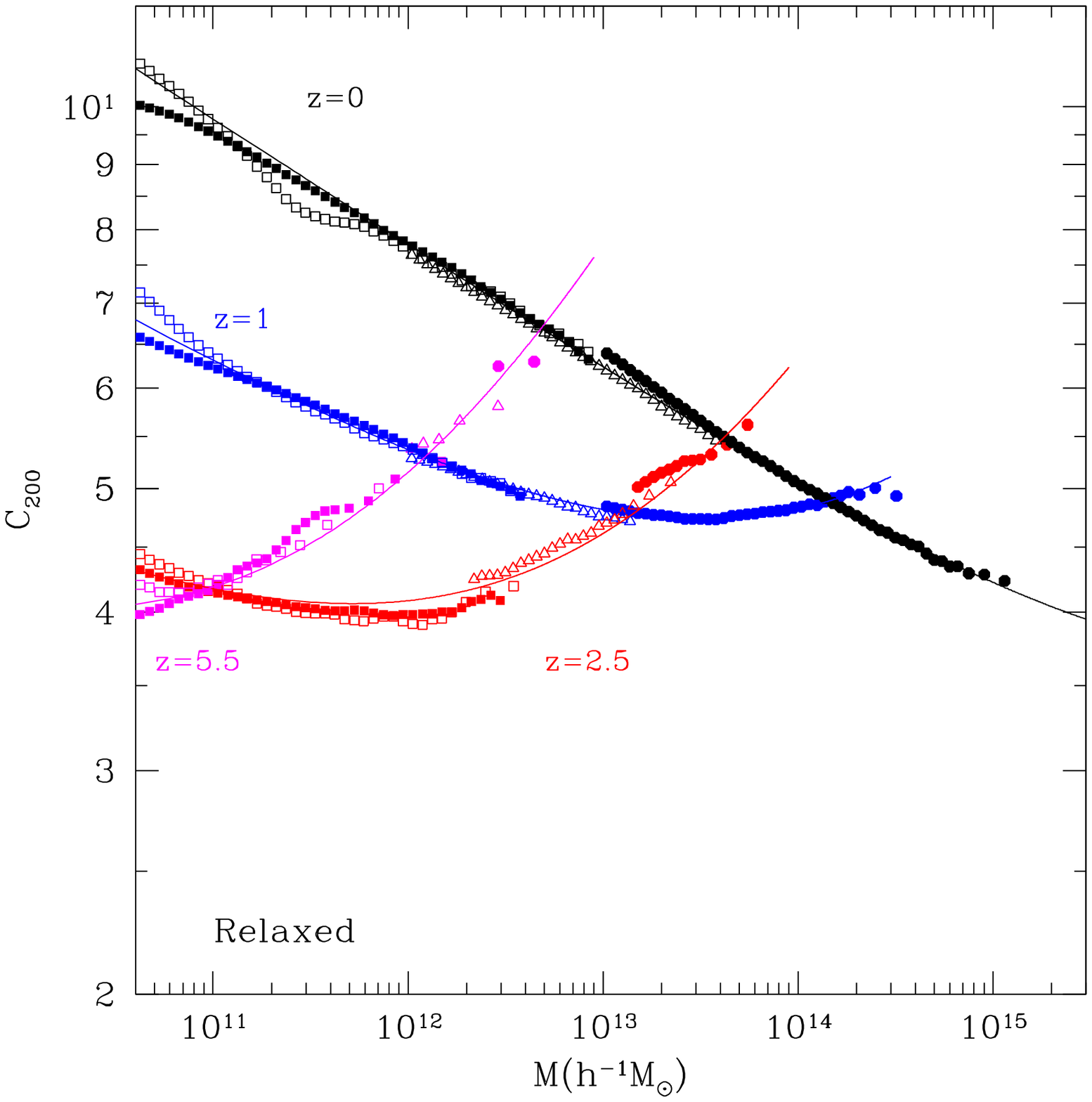}
\caption{Evolution of halo concentration in the
  MultiDark suite of simulations. Shown are the results for simulations
  with the Planck cosmological parameters. The overdensity 200
  criterion was used to find viral mass and radius. {\it Left panel:}
  All halos are binned by the maximum circular velocity
  $\Vmax$. Bottom axis presents corresponding average virial mass of
  these halos.  {\it Right panel:} Relaxed halos are binned by virial
  mass. Full curves in the plots show fits given by
  eq.(\ref{eq:powerfit}).  Relaxed halos have larger concentrations
  and shallower slopes at small masses. Qualitatively the evolution of
  halo concentration does not depend on selection of halos.  }
\label{fig:conc_vmax}
\end{figure*}

In addition to cosmological parameters, redshift, mass, and definition
of virial radius, the halo concentration depends on how halos are
selected. There are a number of options to do that. One can select all halos
or only relaxed halos. Halos can be selected by virial mass or by
circular velocity. Not surprisingly, different selection options
result in a large number of combinations of $C(M)$
approximations. However, {\it qualitatively} the results are the same. 
Figure~\ref{fig:conc_vmax} gives two examples of $C(M,z)$ for halos
selected in different ways.  Here, we show the evolution of halo
concentration defined using eq.~(\ref{eq:Vthree}) for halos selected
by $\Vmax$ and by $M_{200}$. The plots illustrate a well known result
that the evolution of $C(M,z)$ is quite complex \citep[e.g.][]{Prada2012}. Nevertheless,
qualitatively the pattern of the evolution is the same regardless how
halos are selected.

The plots in Figure~\ref{fig:conc_vmax} show that there are three
regimes for $C(M)$: a  declining part on small masses, an upturn at very
large masses and a plateau in between. Curves for $z=1$ and $z=2.5$
clearly illustrate this behavior.  There is no declining part of
$C(M)$ at very large redshifts because our simulations do not have
enough resolution to probe the declining brunch of $C(M,z)$.

As discussed in the previous section, the upturn in the concentration
is related with the increase of $\alpha(M)$ at large masses.  This is
different for halos in the declining regime where the concentration
declines mostly due to the change in the core radius $r_{-2}$ while
$\alpha(M)$ stays nearly constant.

\begin{figure*}
\centering
\includegraphics[width=0.48\textwidth]
{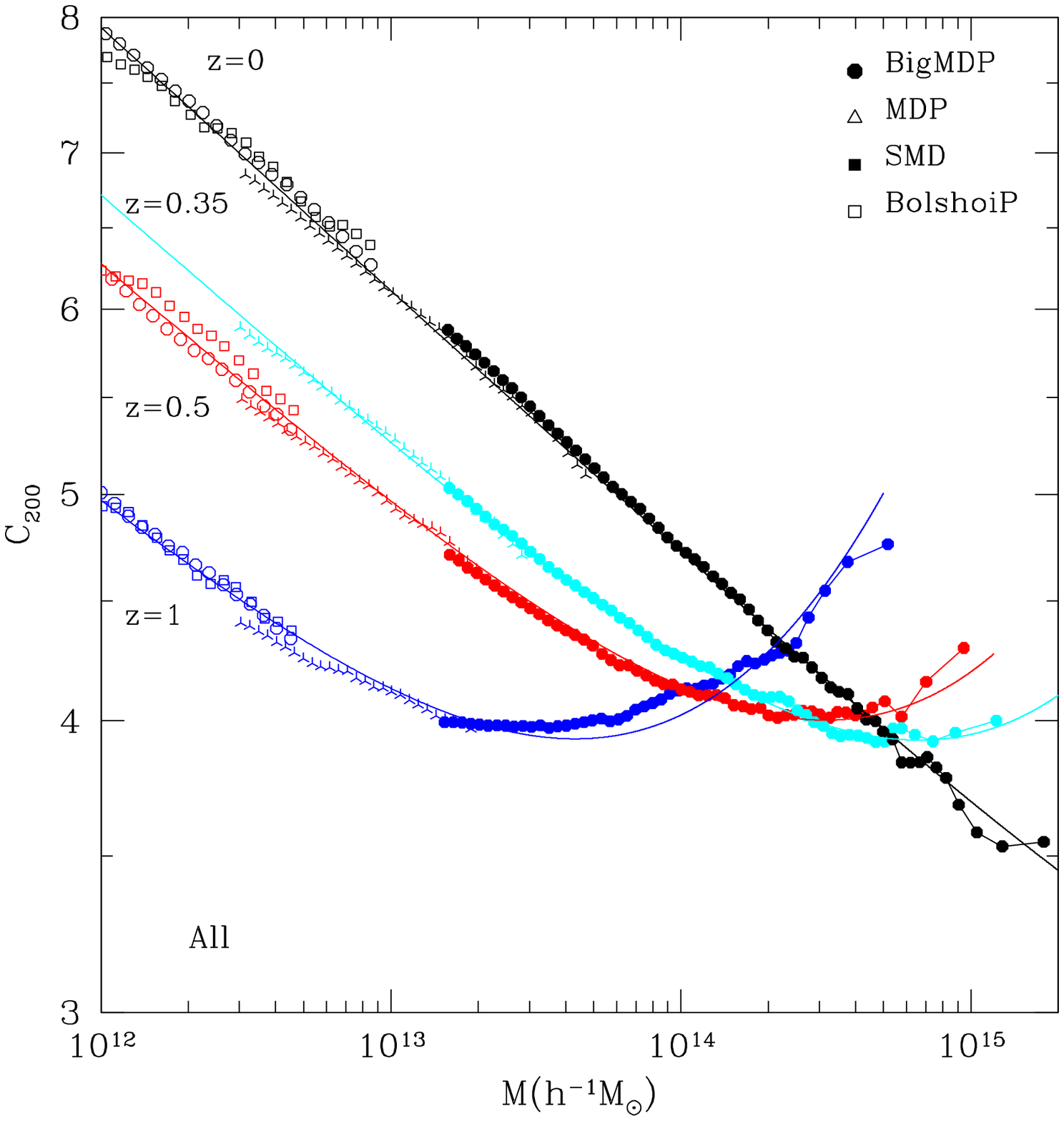}
\includegraphics[width=0.48\textwidth]
{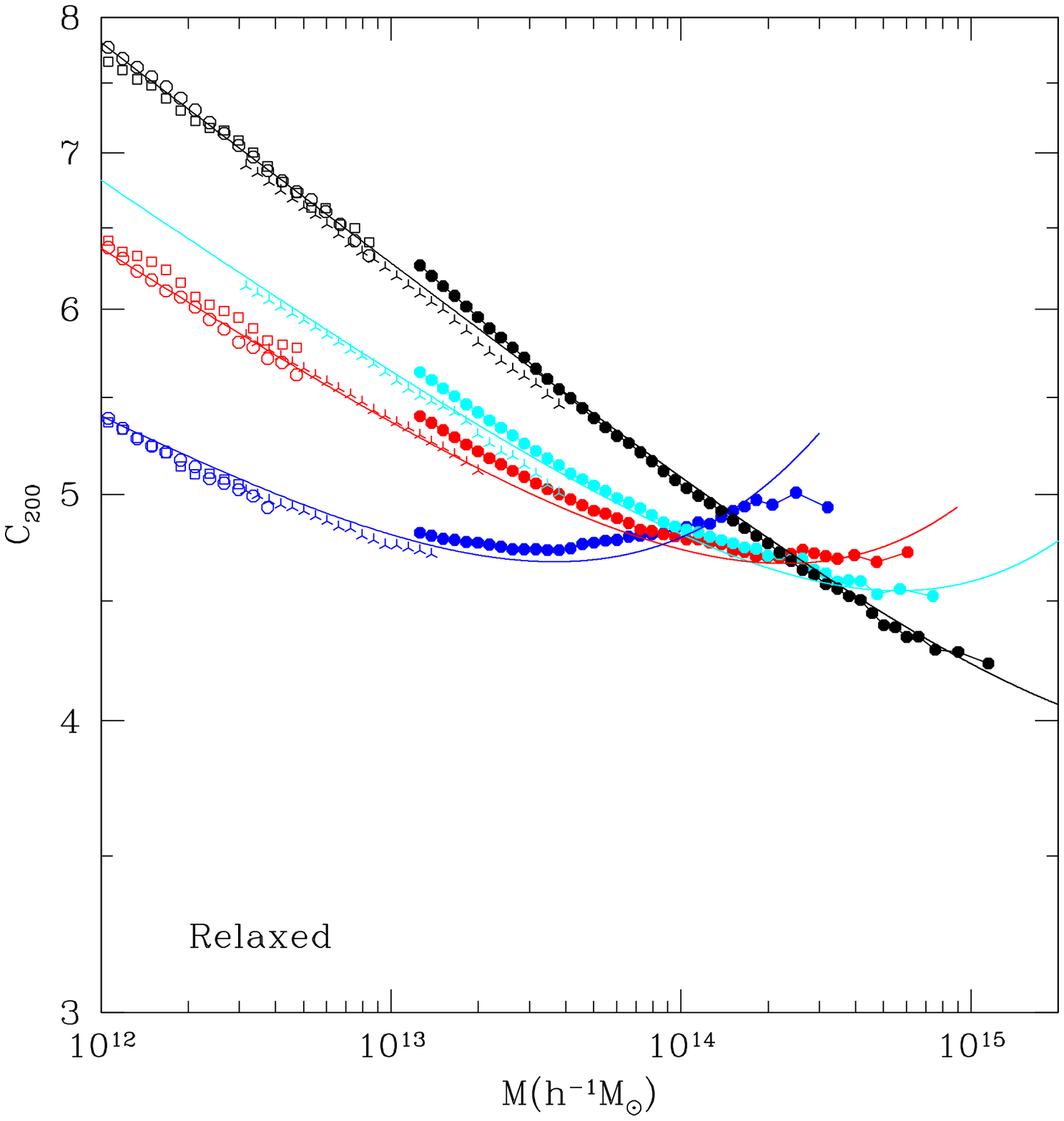}
\caption{The same as in Figure~\ref{fig:conc_vmax} but for low
  redshifts.  In each panel from top to bottom results are presented
  for redshifts $z =0, 0.35, 0.5, 1$. The plot shows how the upturn in
  $c(M)$ seen at high redshifts gradually disappears as the Universe
  becomes dominated by the dark energy. }
\label{fig:LateEvolution}
\end{figure*}

There is no upturn at $z=0$. In order to shed light on this, we
explore the evolution of halo concentration at low redshifts.
Figure~\ref{fig:LateEvolution} shows that the concentration on the
plateau starts to decline at $z\approx 0.5$ and the upturn gets
smaller and disappears at $z=0$. Though the reason is not exactly
clear, we speculate that this likely is related with the decline in
matter density $\Omega_m(z)$ and, consequently,  slower mass
accretion on halos.

\begin{table}
  \caption{Parameters for the concentration - mass relation given by
 eq.(\ref{eq:powerfit}) for the Planck cosmology. Halos are defined using
 the overdensity $200\rho_{\rm cr}$ criterion. All halos are selected by mass. }
\begin{center}
  \tabcolsep 7.2pt
\begin{tabular*}{0.4\textwidth}{@{}lccl@{}}
\hline\hline
 &  \multicolumn{3}{c}{Parameter}\\
Redshift & $C_0$          & $\gamma$ & $M_0/10^{12}\Msunh$  \\
\hline
&&&\\
0.00 & 7.40 &  0.120  & $5.5\times 10^{5}$ \\
0.35 & 6.25 &  0.117  & $1.0\times 10^{5}$ \\
0.50 & 5.65 &  0.115  & $2.0\times 10^{4}$ \\
1.00 & 4.30 &  0.110  & $900$ \\
1.44 & 3.53 &  0.095  & $300$ \\
2.15 & 2.70 &  0.085  & $42$  \\
2.50 & 2.42 &  0.080  & $17$  \\
2.90 & 2.20 &  0.080  & $8.5$ \\
4.10 & 1.92 &  0.080  & $2.0$ \\
5.40 & 1.65 &  0.080  & $0.3$ \\
&&&\\
\hline\hline
\end{tabular*}
\end{center}
\label{tab:fitpars1}
\end{table}

It is often convenient to have simple fits for concentration as
function of mass for some redshifts. To make those fits we use the
following 3-parameter functional form:
\begin{equation}
  C(M) = C_0\left(\frac{M}{10^{12}h^{-1}M_\odot}\right)^{-\gamma} 
           \left[1+ \left(\frac{M}{M_0}\right)^{0.4}\right].
\label{eq:powerfit}
\end{equation}
Table~\ref{tab:fitpars1} gives parameters $C_0$ and $M_0$ for this approximation for
the Planck cosmology. Parameters for additional selection conditions
and for WMAP7 cosmology are presented in the Appendix~A in 
tables~\ref{tab:fitpars}--~\ref{tab:fitpars4}.

\makeatletter{}

\makeatletter{}

\begin{figure*}
\centering
\includegraphics[width=0.48\textwidth]
{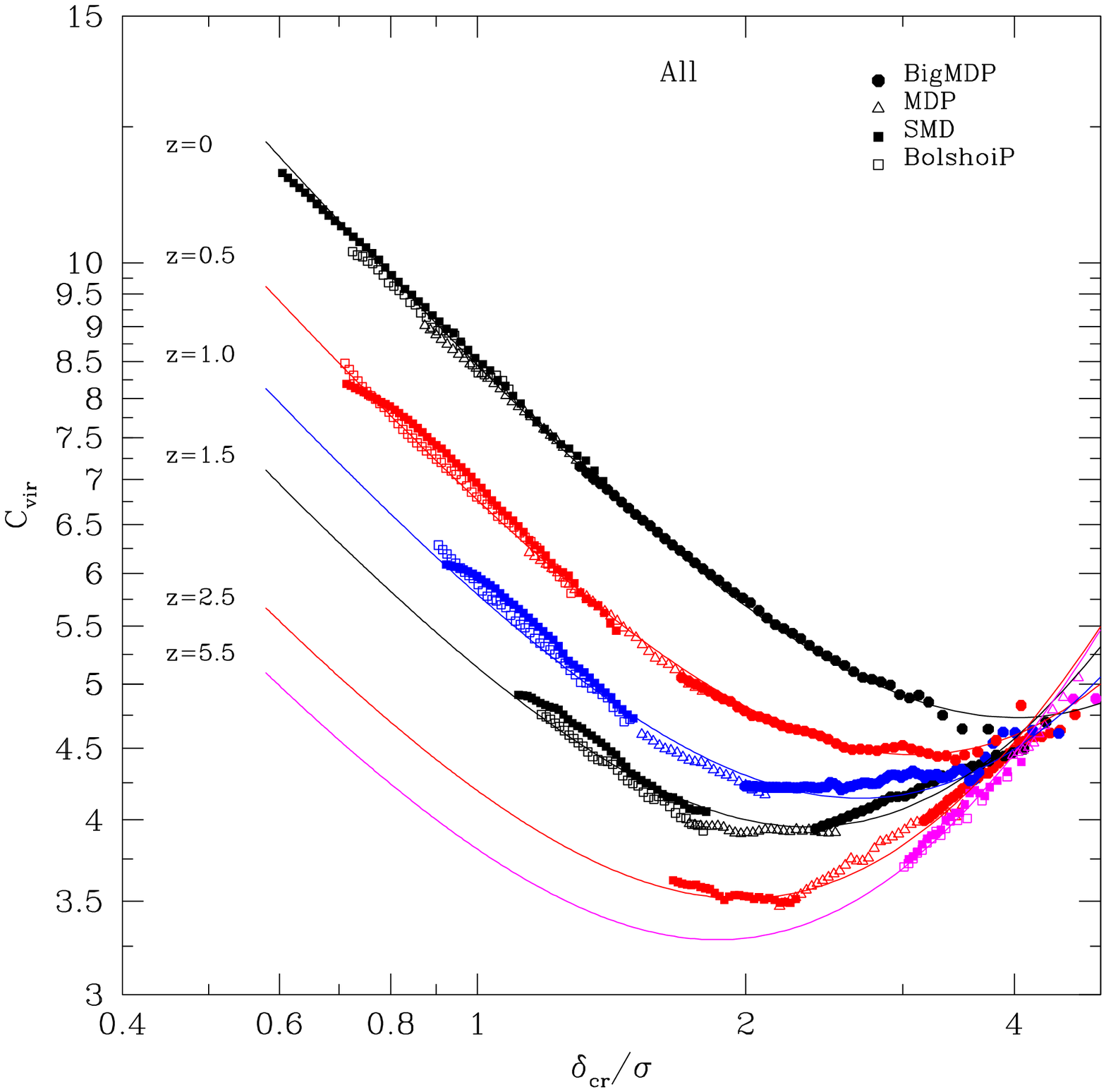}
\includegraphics[width=0.48\textwidth]
{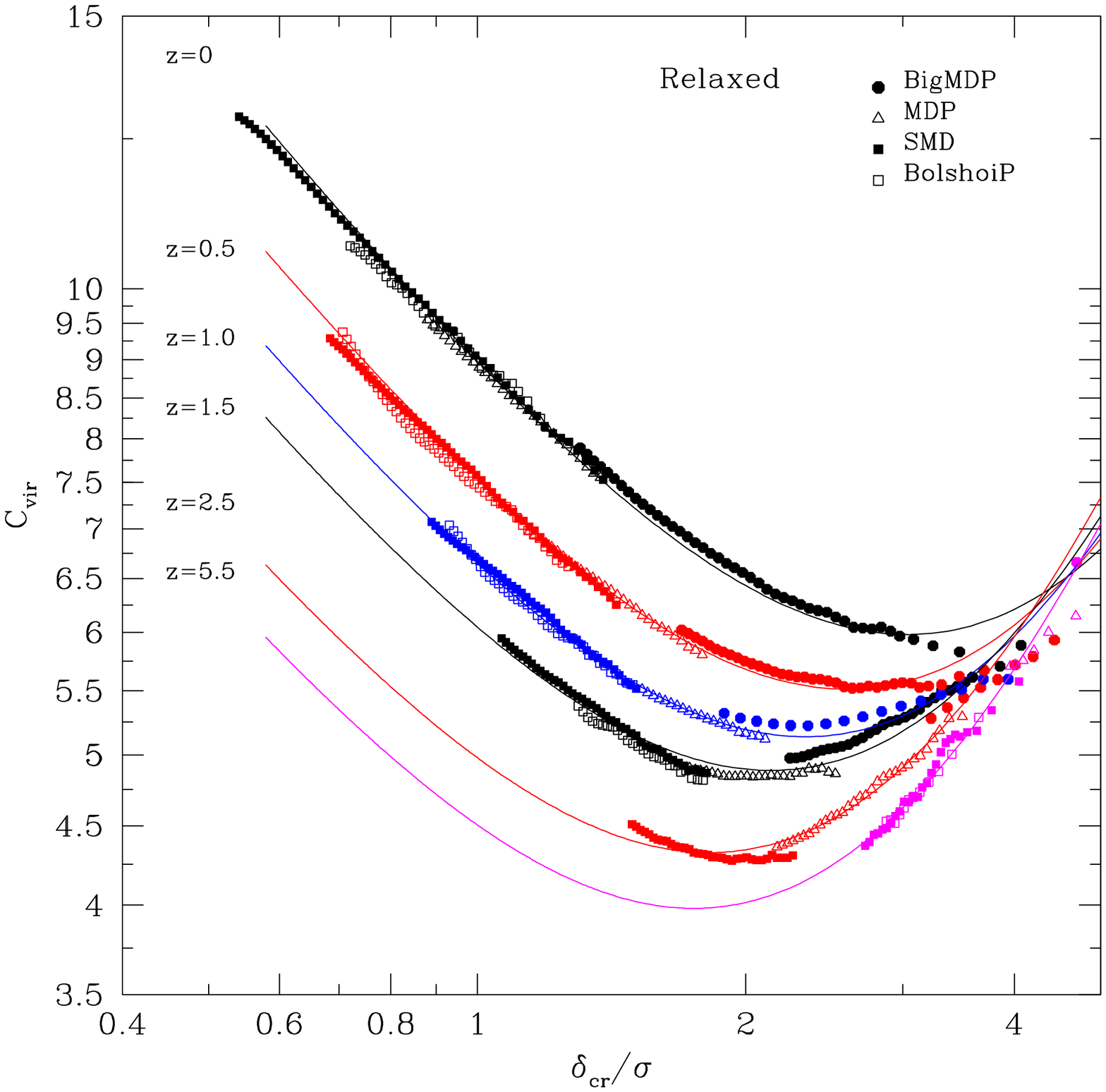}
\caption{Dependence of halo concentration on the peak height $\nu
  =\delta_{\rm cr}/\sigma(M,z)$ for all halos (left panel) and
relaxed halos (right panel). Full curves show analytical fits given by eq.~(\ref{eq:FitNu}).
\label{fig:ConcSigmaMass}}
\end{figure*}

Following \citet{Prada2012} we study the evolution of halo
concentrations as function of the amplitude of perturbations
$\sigma(M)$ or as function of peak height $\nu(M)$.  Evolution of the
$C(\nu)$ relation for halos defined using the virial mass $\Mvir$ and
radius $\Rvir$ is presented in Figure~\ref{fig:ConcSigmaMass} for the
Planck cosmology for all and relaxed halos.  As
Figure~\ref{fig:ConcSigmaMass} shows, the evolution of $C(\nu)$ looks
more simple as compared with $C(M,z)$ . As \citet{Prada2012} and
\citet{DiemerKravtsovC} found, the shape of $C(\nu)$ is almost the
same at every redshift, but its position in the $C-\nu$ diagram
gradually shifts.  Here we find that the following functional form
describes the shape of $C(\nu)$:
\begin{equation}
C(\sigma) = b_0\left[1+7.37\left(\frac{\sigma}{a_0}\right)^{3/4}\right]
 \left[1+0.14\left(\frac{\sigma}{a_0}\right)^{-2}\right]
\label{eq:FitNu}
\end{equation}

\makeatletter{}
 
Parameters $b_0$ and $a_0$ for different redshifts and selection
conditions are given in the Appendix~A in tables~\ref{tab:fitNu200} and ~\ref{tab:fitNu}.
Appendix~B gives examples of evolution of halo concentration for few large halos.

\section{Upturn of halo concentration}
\label{sec:Upturn}
The upturn in the concentration -- mass relation is probably the most
controversial regime.  It was first discovered in \citet{Klypin2011}
in the Bolshoi simulation made with the ART code.  It later was found
in the Millennium GADGET simulations and in the ART Multidark
simulation by \citet{Prada2012} who made extensive analysis of halos
at the upturn. 

Considering that the upturn is a relatively new feature, some
explanations were proposed. \citet{Ludlow2012} argue that the upturn
is an artifact of halos that are out of equilibrium. Their idea can be
summarized as follows. The most massive halos grow fast. When those
halos are caught in the process of first major merger, the infalling
satellite penetrates deep into the major halo producing an illusion of
a very concentrated halo. When \citet{Ludlow2012} removed halos, which
they perceived to be unrelaxed, they find that there is no upturn. The
problem with this idea is that \citet{Ludlow2012} did not make the
pressure term correction to the virial ratio. This renders most of the
most massive halos to be considered unrelaxed when they actually
are. The lack of infalling (radial) velocities in the central region
of $\sim 1/3$ of the virial radius is also not consistent with the
idea of ``first infall'' as an explanation for the upturn. So, the
non-equilibrium explanation for the upturn does not seem to be valid.

Another idea for the upturn is related with the fact that in
\citet{Klypin2011} and \citet{Prada2012} the concentration was
estimated using the $\Vmax/V_{\rm vir}$ ratio.  So, one may speculate
that the upturn is due to the procedure. However, \citet{Prada2012}
and more recently \citet{Dutton2014} compare estimates of
concentration done with fitting the NFW profiles and with the
$\Vmax/V_{\rm vir}$ ratio. They find that both methods produce nearly
identical results for most of halos. Systematic $\sim$10\% differences
in concentration were found only for the most massive halos. However,
the problem with the fitting  densities of the massive halos is that the
profiles substantially deviate from the NFW profile. For these halos
only the ratio $\Rvir/r_{-2}$ was reported as a concentration, thus
missing the main contribution to the concentration, i.e. the increase of
$\alpha$. 

The properties of halos at the upturn point into a different explanation
for the origin of the upturn. These extremely massive halos form around very high
density peaks of initial linear density perturbations. The analysis of
statistical properties of gaussian random fields shows that these rare
peaks tend to be more spherical than low-$\nu$ peaks
\citep{Doroshkevich1970,Bardeen1986}. In turn, this means that infall
velocities are also more radial resulting in deeper penetration of
infalling mass into the halo. This produces more centrally
concentrated halos.

This large concentration will not be preserved as more material is
accreted. As the mass grows, the relative peak height becomes lower and
the infall velocities become less radial. The halo gradually slides
into the plateau regime. As the rate of accretion decreases even more,
the central region stops to be severely affected by mergers, which now
tend to build more extended outer regions. This will be the ``normal''
slow growth mode of halos and associated with the decline regime in the
concentration -- mass relation.

\section{Comparison with other results}
\label{sec:Compare}

There are some differences between our results with those in
\citet{Prada2012}. We now have more data, which allow us to make more
detailed analysis of concentrations. We also make fits of density
profiles using the Einasto profiles. In \citet{Prada2012} we used only
bound particles to estimate profiles and concentrations. Even for
the most massive halos the average fraction of unbound particles is
less than $\sim$3\%. However, this increases the halo concentration by
as much as $\sim 10$\%.
\makeatletter{}\begin{figure*}
\centering
\includegraphics[width=0.47\textwidth]
{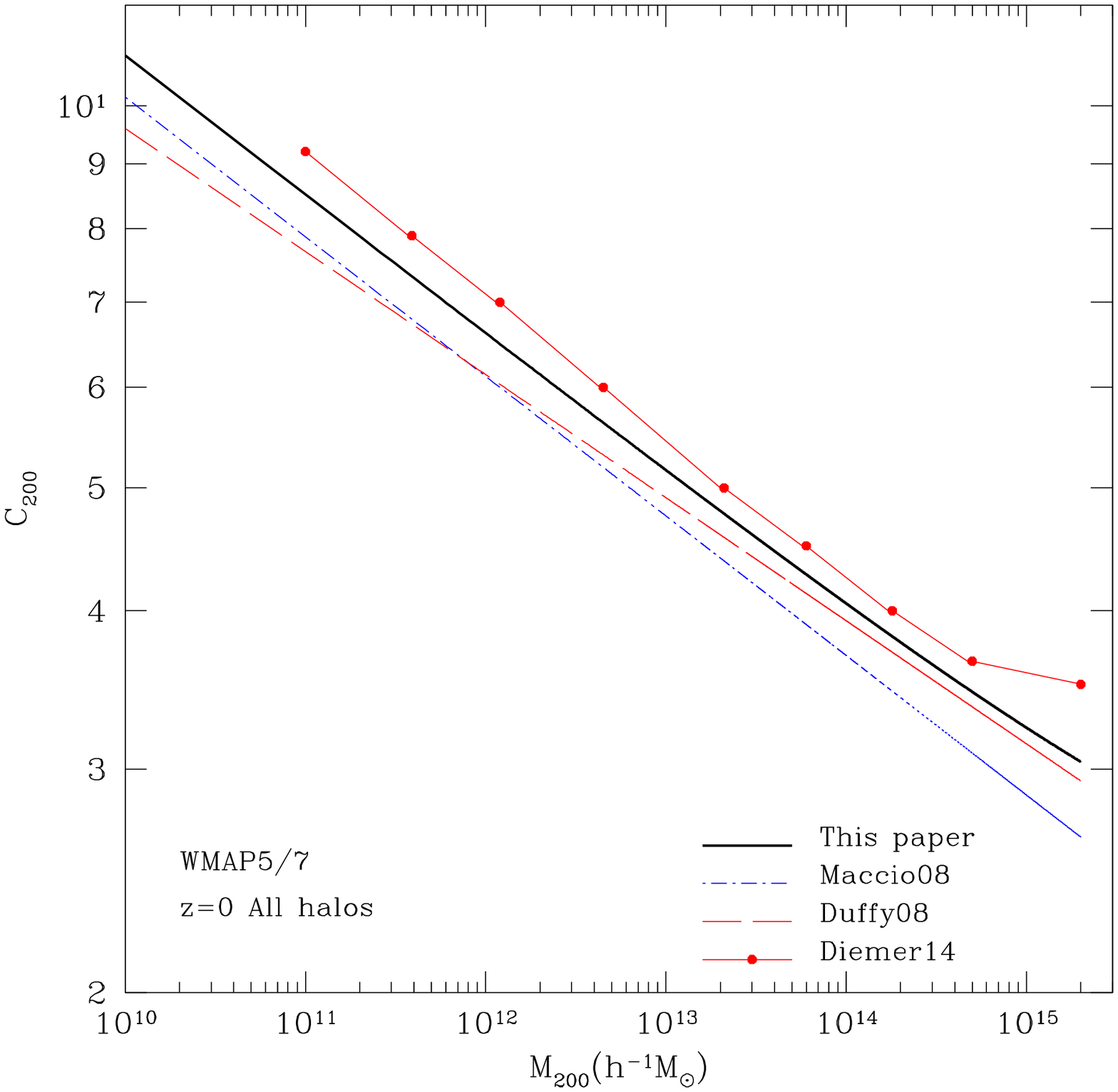}
\includegraphics[width=0.47\textwidth]
{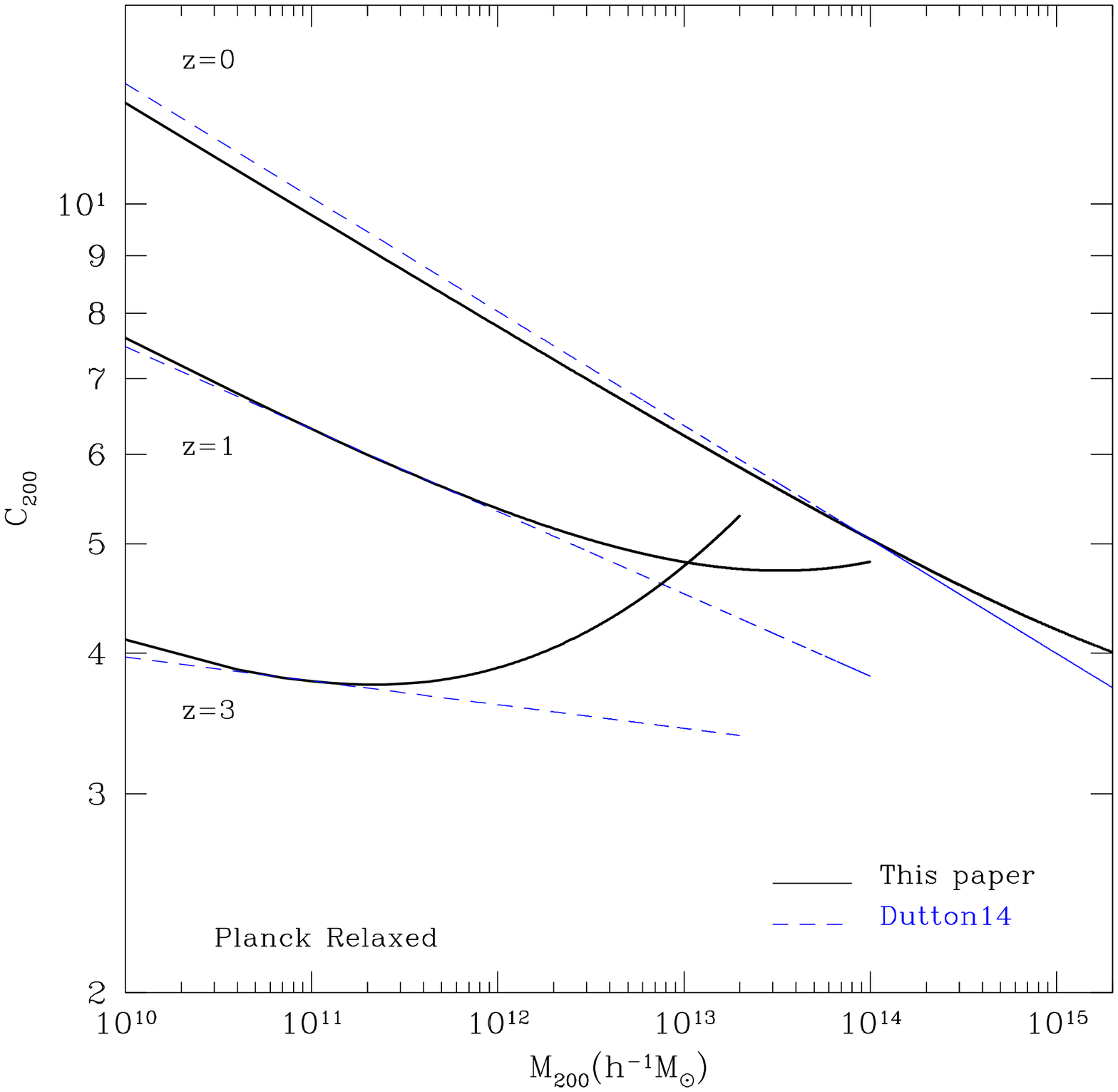}
\caption{Comparison of predictions of dark matter halo concentrations
  defined using the NFW profiles. {\it Left:} Comparison of different
  published results of concentrations for all halos in the WMAP5/7
  cosmologies. Overall the agreement between different groups is quite
  reasonable, but there are some systematic differences. 
  Results of \citet{Duffy2008} and \citet{Maccio2008} are
  systematically below our estimates by $\sim 10$\%. Concentrations in
  \citet{Diemer2014} are $\sim$5\% above ours. {\it Right:} Comparison
  of concentrations of relaxed halos in simulations with the Planck
  cosmology at different redshifts. Agreement between our results and
  those of \citet{Dutton2014} is very good for $z=0$ and for small
  masses at high redshifts. Disagreement at large masses is due to
  fitting  NFW profiles to halos that
  substantially deviate from the NFW shape. Agreement improves once
  the Einasto profiles are used.}
\label{fig:CompareOne}
\end{figure*}

We start with comparing different published estimates of halo
concentrations at redshift $z=0$.  In the left panel of
Figure~\ref{fig:CompareOne} we compare different published results of
concentrations defined using the NFW profiles for all halos in the
WMAP5/7 cosmologies. Overall, the agreement between different groups is
quite reasonable, but there some systematic difference. For example,
results of \citet{Duffy2008} and \citet{Maccio2008} are systematically
below our estimates by $\sim 10$\%. Concentrations in
\citet{Diemer2014} are $\sim$5\% above ours. They also find some
indication of an upturn at the largest masses, which we do not see in
our simulations.
\makeatletter{}\begin{figure*}
\centering
\includegraphics[width=0.47\textwidth]
{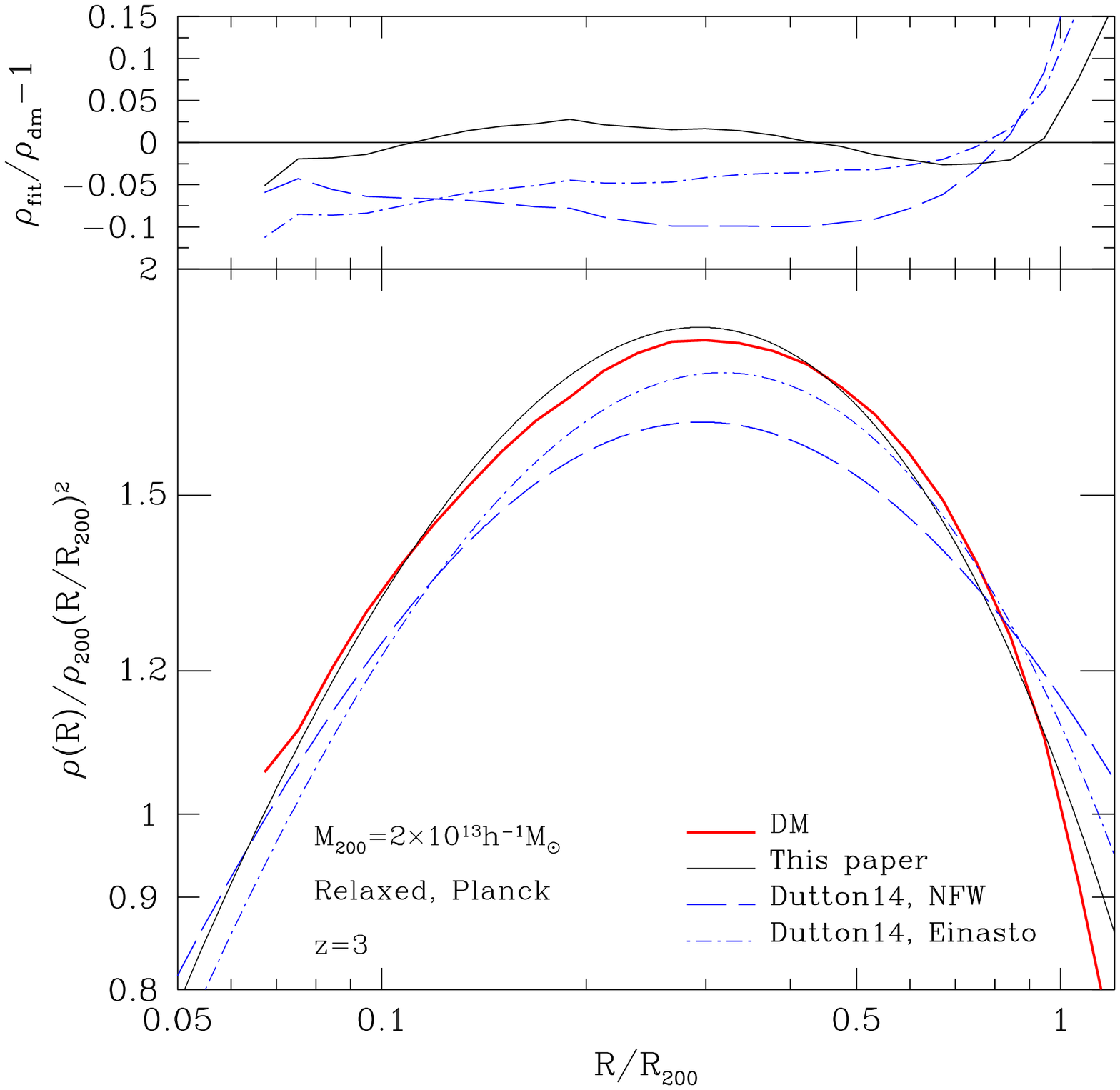}
\includegraphics[width=0.47\textwidth]
{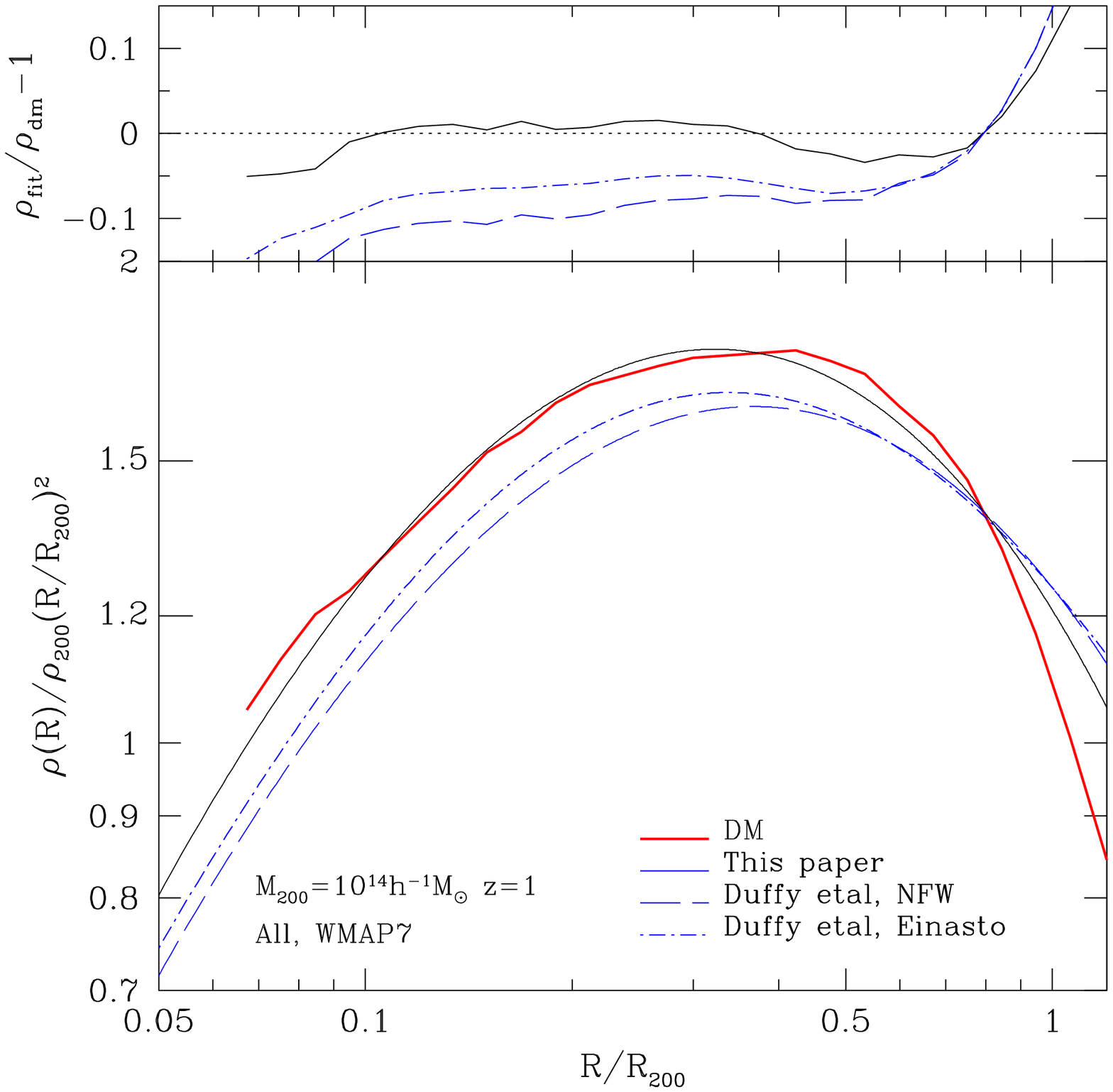}
\caption{Comparison of predictions of halo density profiles at
  different redshifts. Full thick curves in both panels show density
  profiles multiplied by radius squared and normalized to the density
  at $R_{200}$. Top panels show errors of different
  approximations. Approximations are normalized to have the same mass
  as the dark matter halos. {\it Left:} Profiles for relaxed halos in
  the Planck cosmology simulations with mass $\M200=2\cdot
  10^{13}\Msunh$ at $z=3$. Short and long dashed curves show
  predictions for the NFW and Einasto profiles by \citet{Dutton2014}.
}
\label{fig:CompareTwo}
\end{figure*}

We compare concentrations for the Planck cosmology at different
redshifts in the right panel of Figure~\ref{fig:CompareOne}. Here, we
compare NFW-based concentrations for relaxed halos in our simulations
with those in \citet{Dutton2014}. There is an excellent agreement for
$z=0$ with deviations less than $\sim$2\%. The situation is more
complicated at high redshifts with again very good agreement at low
masses and substantial differences at large masses. However, one
should remember that \citet{Dutton2014} fit the NFW profiles to the
data and we do not. Because the most massive halos at high redshifts
have large deviations from the NFW profiles, not surprisingly we have
large differences between our estimates and those of
\citet{Dutton2014}.

Because halo concentrations are used to estimate density profiles, it
is important to compare different predictions for dark matter density
profiles, not just concentrations. We give two examples of such a
comparison. In the right panel of Figure~\ref{fig:CompareTwo} we show
the median density profile of all halos at redshift $z=1$ with mass
$M_{200}=10^{14}\Msunh$ in the WMAP7 MultiDark simulation.  The black
thin curve also shows our approximation using the Einasto profile with
parameters defined by eqs.(\ref{eq:alpha}-\ref{eq:alphaRelax}). We
also compare our results with the predictions given in Table~1 of
\citet{Duffy2008} for both  NFW and Einasto profiles. Those
predictions are systematically below the results of simulations by
$\sim 10$\%. Note that the problem cannot be fixed by scaling up the
density profiles to improve the accuracy of the fits. This cannot be
done because the total halo mass will be substantially larger than the mass
of the halo in the simulation.

The failure of the Einasto fit in the \citet{Duffy2008} approximation
is related to their approximation for the $\alpha(\nu)$
parameter. \citet{Duffy2008} used results from \citet{Gao2008}
(eq.~\ref{eq:alphaDutton}) that gives a $\sim $10\% smaller value of
$\alpha$ as compared with our estimate given by
eq.~(\ref{eq:alpha}). Combined with a slightly smaller value of the
formal concentration, this results in the halo density profile, which
is systematically misses by $\sim$10\% the density in the simulation.

Comparison of our density profiles with those by \citet{Dutton2014}
shows much better agreement as indicated in the left panel of
Figure~\ref{fig:CompareTwo} when the Einasto profile is used. Here, we
show density profile for relaxed halos with $M_{200}=2\cdot
10^{14}\Msunh$ in the Planck cosmology simulation MDPL at $z=3$. The
total of $\sim$300 halos where used to produce the median
profile. Halos of this mass and redshift have $\nu=3.9$, which puts
them in the upturn of the concentration - mass relation.  Our
analytical fit for the density profile is obtained using
eq.~(\ref{eq:alphaRelax}) for the $\alpha$ parameter. The
$R_{200}/r_{-2}$ ratio is estimated using eq.~(\ref{eq:cc}), which
we scale up by factor 1.1 to account for the fact that relaxed halos
are more concentrated than all halos. To construct profiles predicted
by \citet{Dutton2014}, we use parameters for the NFW and Einasto
approximations from their Table~3. All
analytical fits are normalized to have the same total halo mass.

It is clear that the NFW fit given by \citet{Dutton2014} provides a
halo, which is less concentrated than that in the simulation. This is
consistent with the differences which we saw in the right panel of
Figure~\ref{fig:CompareOne}: the NFW fits predict too low
concentrations for halos at the upturn resulting in $\sim 10$\% too
low densities.
However, the differences between our and \citet{Dutton2014} Einasto
fits are significantly smaller. 

\section{How to estimate density profiles}
We offer different options to estimate density profiles of dark matter
halos.  Halos in the declining branch of the $C(M,z)$ relation ($\nu\lesssim 2$) can be
approximated using the NFW profile. Tables in the Appendix~A give
numerous parameters for halos selected in different ways. One can
interpolate between the values provided by the Tables to find the 
approximations for different redshifts. The interpolation of $C(\nu)$
parameters is expected to be more accurate because there is less
evolution of the parameters with the redshift.

There are two options for handling halos at the plateau and upturn.
For all halos in the Planck cosmology one can use
eqs.(\ref{eq:alpha}-\ref{eq:cc}) to find $\alpha$ and
$R_{200}/r_{-2}$. For other selection conditions and for the  WMAP
cosmology, one should use eqs.(\ref{eq:alpha}) or
eqs.(\ref{eq:alphaRelax}) to find $\alpha$. These relations are
expected to be least dependent on details of selection and
cosmological parameters. Then one should use one of the approximations
for $C(\sigma)$ given in the Appendix A. Finally, solve
eqs.(\ref{eq:Vtwo}-\ref{eq:Vfour}) to find $C_E \equiv R_{200}/r_{-2}$ for known $C$ and
$\alpha$.

\section{Conclusions}

In this paper we present the new suite of MultiDark cosmological
N-body simulations from which we have identified more than 60 billion
dark matter halos with more than 100 particles that span more than 5
orders of magnitude in mass and covers more than 50 Gpc$^3$ in volume.
From this large data set we have studied with very high accuracy the
halo density, infall velocity and velocity anisotropy profiles and
concentrations in three dynamical regimes: declining concentration,
plateau, and upturn.  We derive analytical approximations that provide
2--5\% accurate estimates for halo concentrations and density
profiles.

We summarize  the main results from this work:

 \begin{itemize}

 \item In order to understand the evolution of halo concentration and,
   specifically the nature of the upturn, one needs to realize that
   the halo concentration is not defined as the ratio of the virial
   radius to the radius $r_{-2}$ as in the NFW profile. For massive
   halos the average density profile is far from the NFW shape and the
   concentration is not defined by the core radius $r^{-2}$.  In
   Section~5 we present density profiles that clearly show that
   massive halos at $z=3$ have increasing concentrations with
   increasing mass and they have nearly unchanging
   $\Rvir/r_{-2}$. Both parameters $\alpha$ and $r_{-2}$ of the
   Einasto approximation affect the concentration.

 \item We speculate that the increase in the halo concentrations for
   the most massive halos is related with the tendency of rare peaks
   in the random gaussian linear density field to be more spherical
   \citep{Doroshkevich1970,Bardeen1986}. Very radial accretion onto
   these peaks is clearly seen in Figure~\ref{fig:Beta} with the
   average velocity anisotropy parameter $\beta=0.5-0.7$ for halos
   with $\nu\approx 4$. This radial infall brings mass closer to the
   center, producing a highly concentrated halo. As time goes on the
   halo slides into the plateau regime, and accretion becomes less
   radial. Now mass is deposited at larger radius, and the
   concentration declines. Once the rate of accretion and merging
   slows down, the halo moves into the domain of declining $C(M)$
   because new accretion piles up mass close to the virial radius
   while the core radius is staying constant.

 \item The ratio of the maximum circular velocity to the virial
   velocity $\Vmax/V_{\rm vir}$ gives a profile-independent measure of
   the halo concentration.

\item Density profiles of very massive halos with $\nu>3$
  substantially deviate from the NFW shape. Fitting an NFW profile for
  these halos gives incorrect results regardless on how the fitting
  (what range of radii) is done. The differences between the NFW and
  Einasto concentrations for these massive halos do not mean that
  there are uncertainties in halo concentration. They simply indicate
  that the NFW formula should not be used.

\item Differences between the $\Vmax/V_{\rm vir}$ concentration and
  the formal concentration obtained by fitting the Einasto profile
  (the ratio $\Rvir/r_{-2}$) do not indicate that there are real
  disagreements. These two estimates are for different quantities. The
  $\Rvir/r_{-2}$ ratio is only a part of the real halo concentration.
\end{itemize}

\section*{Acknowledgements}
The BigMultidark simulations have been performed on the SuperMUC
supercomputer at the Leibniz-Rechenzentrum (LRZ) in Munich, using the
computing resources awarded to the PRACE project number
2012060963. The SMDPL and HMDPL simulations have been performed on
SuperMUC at LRZ in Munich within the pr87yi project.  Bolshoi(P) and
MultiDark simulations were performed on Pleiades supercomputer at the
NASA Ames supercomputer center.  S.~H. wants to thank R.~Wojtak for
useful discussions. The authors want to thank V.~Springel for
providing us with the optimized version of {\small GADGET-2}.
S.H. acknowledges support by the Deutsche Forschungsgemeinschaft under
the grant $\mathrm{GO}563/21-1$.  A.K. acknowledges support of NSF
grants to NMSU.  GY acknowledges support from MINECO (Spain) under
research grants AYA2012-31101 and FPA2012-34694 and Consolider Ingenio
SyeC CSD2007-0050 FP acknowledge support from the Spanish MICINNs
Consolider-Ingenio 2010 Programme under grant MultiDark CSD2009-00064,
MINECO Centro de Excelencia Severo Ochoa Programme under grant
SEV-2012-0249, and MINECO grant AYA2014-60641-C2-1-P

\bibliography{mdark}
\bibliographystyle{mn2e.bst}

\appendix
\section{Parameters for the concentration - mass relation}

Tables~\ref{tab:fitpars} -- \ref{tab:fitNu} give parameters for approximation eq.(\ref{eq:powerfit}) for
different simulations, virial mass definitions, halo selection criteria, and redshifts.

\begin{table}
  \caption{Parameters for the concentration - mass relation given by 
     eq.(\ref{eq:powerfit}) for Planck cosmology. Halos are defined
     using the overdensity 200 criterion. }
\begin{center}
  \tabcolsep 7.2pt
\begin{tabular*}{0.48\textwidth}{@{}lccl@{}}
\hline\hline
 &  \multicolumn{3}{c}{Parameter}\\
Redshift & $C_0$          & $\gamma$ & $M_0/10^{12}\Msunh$  \\
\hline
 &\multicolumn{2}{l}{Relaxed halos selected by mass}\\
0.00 & 7.75 &  0.100  & $4.5\times 10^{5}$ \\
0.35 & 6.70 &  0.095  & $2.0\times 10^{4}$ \\
0.50 & 6.25 &  0.092  & $8.0\times 10^{3}$ \\
1.00 & 5.02 &  0.088  & $780$ \\
1.44 & 4.19 &  0.085  & $160$ \\
2.15 & 3.30 &  0.083  & $27$  \\
2.50 & 3.00 &  0.080  & $14$  \\
2.90 & 2.72 &  0.080  & $6.8$ \\
4.10 & 2.40 &  0.080  & $1.6$ \\
5.40 & 2.10 &  0.080  & $0.30$ \\
&&&\\
      &\multicolumn{2}{l}{All halos selected by mass}\\
0.00 & 7.40 &  0.120  & $5.5\times 10^{5}$ \\
0.35 & 6.25 &  0.117  & $1.0\times 10^{5}$ \\
0.50 & 5.65 &  0.115  & $2.0\times 10^{4}$ \\
1.00 & 4.30 &  0.110  & $900$ \\
1.44 & 3.53 &  0.095  & $300$ \\
2.15 & 2.70 &  0.085  & $42$  \\
2.50 & 2.42 &  0.080  & $17$  \\
2.90 & 2.20 &  0.080  & $8.5$ \\
4.10 & 1.92 &  0.080  & $2.0$ \\
5.40 & 1.65 &  0.080  & $0.3$ \\
&&&\\
\hline\hline
\end{tabular*}
\end{center}
\label{tab:fitpars}
\end{table}

\begin{table}
  \caption{Parameters for the concentration - mass relation given by 
     eq.(\ref{eq:powerfit}) for Planck cosmology. Halos are defined
     using the overdensity 200 criterion. }
\begin{center}
  \tabcolsep 7.2pt
\begin{tabular*}{0.48\textwidth}{@{}lccl@{}}
\hline\hline
 &  \multicolumn{3}{c}{Parameter}\\
Redshift & $C_0$          & $\gamma$ & $M_0/10^{12}\Msunh$  \\
\hline
      &\multicolumn{2}{l}{Relaxed halos selected by $\Vmax$}\\
0.00 & 8.0 &  0.100  & $2.0\times 10^{5}$ \\
0.35 & 6.82 &  0.095  & $9.0\times 10^{3}$ \\
0.50 & 6.40 &  0.092  & $4.5\times 10^{3}$ \\
1.00 & 5.20 &  0.088  & $600$ \\
1.44 & 4.35 &  0.085  & $150$ \\
2.15 & 3.50 &  0.080  & $27$  \\
2.50 & 3.12 &  0.080  & $11$  \\
2.90 & 2.85 &  0.080  & $5.5$ \\
4.10 & 2.55 &  0.080  & $1.5$ \\
5.40 & 2.16 &  0.080  & $0.22$ \\
&&&\\
      &\multicolumn{2}{l}{All halos selected by $\Vmax$}\\
0.00 & 7.75 &  0.115  & $5.5\times 10^{5}$ \\
0.35 & 6.50 &  0.115  & $1.8\times 10^{4}$ \\
0.50 & 5.95 &  0.115  & $6.0\times 10^{3}$ \\
1.00 & 4.55 &  0.110  & $600$ \\
1.44 & 3.68 &  0.105  & $150$ \\
2.15 & 2.75 &  0.100  & $20$  \\
2.50 & 2.50 &  0.095  & $10$  \\
2.90 & 2.25 &  0.090  & $5.0$   \\
4.10 & 2.05 &  0.080  & $1.5$ \\
5.40 & 1.76 &  0.080  & $0.25$ \\
&&&\\
\hline\hline
\end{tabular*}
\end{center}
\label{tab:fitpars0}
\end{table}

\begin{table}
  \caption{Parameters for the concentration - mass relation given by eq.(\ref{eq:powerfit})
   for Planck cosmology. Halos are defined using the virial overdensity criterion. }
\begin{center}
  \tabcolsep 7.2pt
\begin{tabular*}{0.48\textwidth}{@{}lccl@{}}
\hline\hline
 &  \multicolumn{2}{c}{Parameter}\\
Redshift & $c_0$          & $\gamma$ & $M_0/10^{12}\Msunh$ \\
\hline
  &\multicolumn{2}{l}{Relaxed halos selected by mass}\\
0.00 & 10.2 &  0.100  & $1.\times 10^{5}$  \\
0.35 & 7.85  &  0.095  & $1.2\times 10^{4}$  \\
0.50 & 7.16 &  0.092  & $5.5\times 10^{3}$  \\
1.00 & 5.45 &  0.088  & $700$  \\
1.44 & 4.55 &  0.085  & $180$  \\
2.15 & 3.55 &  0.080  & $30$  \\
2.50 & 3.24 &  0.080  & $15$  \\
2.90 & 2.92 &  0.080  & $7.0$  \\
4.10 & 2.60 &  0.080  & $1.9$  \\
5.40 & 2.30 &  0.080  & $0.36$  \\
&&&\\
      &\multicolumn{2}{l}{All halos selected by mass}\\
0.00 & 9.75 &  0.110  & $5.0\times 10^{5}$  \\
0.35 & 7.25 &  0.107  & $2.2\times 10^{4}$  \\
0.50 & 6.50 &  0.105  & $1.0\times 10^{4}$  \\
1.00 & 4.75 &  0.100  & $1000$  \\
1.44 & 3.80 &  0.095  & $210$  \\
2.15 & 3.00 &  0.085  & $43$  \\
2.50 & 2.65 &  0.080  & $18$  \\
2.90 & 2.42 &  0.080  & $9.0$  \\
4.10 & 2.10 &  0.080  & $1.9$  \\
5.40 & 1.86 &  0.080  & $0.42$  \\
&&&\\
\hline\hline
\end{tabular*}
\end{center}
\label{tab:fitpars2}
\end{table}

\begin{table}
  \caption{Parameters for the concentration - mass relation given by eq.(\ref{eq:powerfit})
 for Planck cosmology. Halos are defined using the virial overdensity criterion. }
\begin{center}
  \tabcolsep 7.2pt
\begin{tabular*}{0.48\textwidth}{@{}lccl@{}}
\hline\hline
 &  \multicolumn{2}{c}{Parameter}\\
Redshift & $c_0$          & $\gamma$ & $M_0/10^{12}\Msunh$ \\
\hline
      &\multicolumn{2}{l}{Relaxed halos selected by $\Vmax$}\\
0.00 & 10.7 &  0.110  & $2.4\times 10^{4}$  \\
0.35 & 8.1  &  0.100  & $5.0\times 10^{3}$  \\
0.50 & 7.33 &  0.100  & $2.2\times 10^{3}$  \\
1.00 & 5.65 &  0.088  & $520$  \\
1.44 & 4.65 &  0.085  & $120$  \\
2.15 & 3.70 &  0.080  & $25$  \\
2.50 & 2.35 &  0.080  & $12$  \\
2.90 & 2.98 &  0.080  & $5.0$  \\
4.10 & 2.70 &  0.080  & $1.4$  \\
5.40 & 2.35 &  0.080  & $0.26$  \\
&&&\\
      &\multicolumn{2}{l}{All halos selected by $\Vmax$}\\
0.00 & 10.3 &  0.115  & $4.8\times 10^{4}$  \\
0.35 & 7.6 &  0.115  & $5.0\times 10^{3}$  \\
0.50 & 6.83 &  0.115  & $2.7\times 10^{3}$  \\
1.00 & 4.96 &  0.110  & $390$  \\
1.44 & 3.96 &  0.105  & $110$  \\
2.15 & 3.00 &  0.100  & $18$  \\
2.50 & 2.73 &  0.095  & $10$  \\
2.90 & 2.45 &  0.090  & $5.0$  \\
4.10 & 2.24 &  0.080  & $1.4$  \\
5.40 & 2.03 &  0.080  & $0.36$  \\
&&&\\
\hline\hline
\end{tabular*}
\end{center}
\label{tab:fitpars20}
\end{table}

\begin{table}
  \caption{Parameters for the concentration - mass relation given by eq.(\ref{eq:powerfit}) 
    for WMAP7 cosmology. Halos are defined using the 200 overdensity criterion. }
\begin{center}
  \tabcolsep 7.2pt
\begin{tabular*}{0.48\textwidth}{@{}lccl@{}}
\hline\hline
 &  \multicolumn{2}{c}{Parameter}\\
Redshift & $c_0$          & $\gamma$ & $M_0/10^{12}\Msunh$  \\
\hline
  &\multicolumn{2}{l}{Relaxed halos selected by mass}\\
0.0 & 6.90  &  0.090  & $5.5\times 10^{5}$  \\
0.50 & 5.70  &  0.088  & $6000$  \\
1.00 & 4.55  &  0.086  & $500$  \\
1.44 & 3.75 &  0.085  & $100$  \\
2.15 & 2.9 &  0.085  & $20$  \\
2.50 & 2.6 &  0.085  & $10$  \\
2.90 & 2.4 &  0.085  & $6.0$  \\
4.10 & 2.2 &  0.085  & $3.0$  \\
&&&\\
      &\multicolumn{2}{l}{All halos selected by mass}\\
0.0 & 6.60  &  0.110  & $2\times 10^{6}$  \\
0.50 & 5.25  &  0.105  & $6\times 10^{4}$  \\
1.00 & 3.85  &  0.103  & $800$  \\
1.44 & 3.0 &  0.097  & $110$  \\
2.15 & 2.1 &  0.095  & $13$  \\
2.50 & 1.8 &  0.095  & $6.0$  \\
2.90 & 1.6 &  0.095  & $3.0$  \\
4.10 & 1.4 &  0.095  & $1.0$  \\
&&&\\
      &\multicolumn{2}{l}{Relaxed halos selected by $\Vmax$}\\
0.0 & 7.20  &  0.090  & $2.0\times 10^{5}$  \\
0.50 & 5.90  &  0.088  & $4000$  \\
1.00 & 4.70  &  0.086  & $400$  \\
1.44 & 3.85 &  0.085  & $80$  \\
2.15 & 3.0 &  0.085  & $13$  \\
2.50 & 2.7 &  0.085  & $7.0$  \\
2.90 & 2.5 &  0.085  & $3.5$  \\
4.10 & 2.3 &  0.085  & $2.0$  \\

\hline\hline
\end{tabular*}
\end{center}
\label{tab:fitpars3}
\end{table}

\begin{table}
  \caption{Parameters for the concentration - mass relation given by eq.(\ref{eq:powerfit}) 
    for WMAP7 cosmology. Halos are defined using the virial overdensity criterion. }
\begin{center}
  \tabcolsep 7.2pt
\begin{tabular*}{0.48\textwidth}{@{}lccl@{}}
\hline\hline
 &  \multicolumn{2}{c}{Parameter}\\
Redshift & $c_0$          & $\gamma$ & $M_0/10^{12}\Msunh$  \\
\hline
  &\multicolumn{2}{l}{Relaxed halos selected by mass}\\
0.0 & 9.50  &  0.090  & $3.0\times 10^{5}$  \\
0.50 & 6.75  &  0.088  & $5000$  \\
1.00 & 5.00  &  0.086  & $450$  \\
1.44 & 4.05 &  0.085  & $90$  \\
2.15 & 3.10 &  0.085  & $15$  \\
2.50 & 2.80 &  0.085  & $8.0$  \\
2.90 & 2.45 &  0.085  & $3.5$  \\
4.10 & 2.20 &  0.085  & $1.5$  \\
&&&\\
      &\multicolumn{2}{l}{All halos selected by mass}\\
0.0 & 9.00  &  0.100  & $2\times 10^{6}$  \\
0.50 & 6.00 &  0.100  & $7\times 10^{3}$  \\
1.00 & 4.30 &  0.100  & $550$  \\
1.44 & 3.30 &  0.100  & $90$  \\
2.15 & 2.30 &  0.095  & $11$  \\
2.50 & 2.10 &  0.095  & $6.0$  \\
2.90 & 1.85 &  0.095  & $2.5$  \\
4.10 & 1.70 &  0.095  & $1.0$  \\
&&&\\
      &\multicolumn{2}{l}{Relaxed halos selected by $\Vmax$}\\
0.0 & 9.75  &  0.085  & $1.3\times 10^{5}$  \\
0.50 & 7.02  &  0.085  & $4000$  \\
1.00 & 5.23  &  0.085  & $400$  \\
1.44 & 4.25 &  0.085  & $80$  \\
2.15 & 3.20 &  0.085  & $11$  \\
2.50 & 2.90 &  0.085  & $6.0$  \\
2.90 & 2.50 &  0.085  & $2.5$  \\
4.10 & 2.35 &  0.085  & $1.2$  \\
\hline\hline
\end{tabular*}
\end{center}
\label{tab:fitpars4}
\end{table}

\begin{table}
  \caption{Parameters for the concentration - mass relation given by eq.(\ref{eq:FitNu}) 
    for Planck cosmology. Halos are defined using the 200 critical overdensity criterion. }
\begin{center}
  \tabcolsep 7.2pt
\begin{tabular*}{0.40\textwidth}{@{}lcc@{}}
\hline\hline
 &  \multicolumn{2}{c}{Parameter}\\
Redshift & $b_0$ & $a_0$   \\
\hline
  &\multicolumn{1}{l}{All halos selected by mass}\\
0.0  & 0.278  &  0.40 \\
0.38 & 0.375  &  0.65 \\
0.50 & 0.411  &  0.82 \\
1.00 & 0.436  &  1.08 \\
1.44 & 0.426  &  1.23 \\
2.50 & 0.375  &  1.60 \\
2.89 & 0.360  &  1.68 \\
5.41 & 0.351  &  1.70 \\
&&\\
  &\multicolumn{1}{l}{Relaxed halos selected by mass}\\
0.0  & 0.522  &  0.95 \\
0.38 & 0.550  &  1.06 \\
0.50 & 0.562  &  1.15 \\
1.00 & 0.562  &  1.28 \\
1.44 & 0.541  &  1.39 \\
2.50 & 0.480  &  1.66 \\
2.89 & 0.464  &  1.70 \\
5.41 & 0.450  &  1.72 \\
  &\multicolumn{1}{l}{All halos selected by $\Vmax$}\\
0.0  & 0.41  &  0.65 \\
1.00 & 0.49  &  1.15 \\
2.50 & 0.41  &  1.60 \\
\hline\hline
\end{tabular*}
\end{center}
\label{tab:fitNu200}
\end{table}

\begin{table}
  \caption{Parameters for the concentration - mass relation given by eq.(\ref{eq:FitNu}) 
    for Planck cosmology. Halos are defined using the virial overdensity criterion. }
\begin{center}
  \tabcolsep 7.2pt
\begin{tabular*}{0.40\textwidth}{@{}lcc@{}}
\hline\hline
 &  \multicolumn{2}{c}{Parameter}\\
Redshift & $b_0$ & $a_0$   \\
\hline
  &\multicolumn{1}{l}{All halos selected by mass}\\
0.0  & 0.567  &  0.75 \\
0.38 & 0.541  &  0.90 \\
0.50 & 0.529  &  0.97 \\
1.00 & 0.496  &  1.12 \\
1.44 & 0.474  &  1.28 \\
2.50 & 0.421  &  1.52 \\
5.50 & 0.393  &  1.62 \\
&&\\
  &\multicolumn{1}{l}{Relaxed halos selected by mass}\\
0.0  & 0.716  &  0.99 \\
0.38 & 0.673  &  1.10 \\
0.50 & 0.660  &  1.16 \\
1.00 & 0.615  &  1.29 \\
1.44 & 0.585  &  1.41 \\
2.50 & 0.518  &  1.65 \\
5.50 & 0.476  &  1.72 \\
\hline\hline
\end{tabular*}
\end{center}
\label{tab:fitNu}
\end{table}

\newpage

\section{Examples of evolution of halo concentration and mass accretion history. }
In Sections~\ref{sec:Devolve} and~\ref{sec:Cevolve} we discussed the
evolution of the average properties of halos, including density
profiles and halo concentrations. In this appendix we focus on a
somewhat different but related issue: how individual halos evolve.
The evolution of individual halos has been extensively studied over
the past 15 years
\citep[e.g.,][]{Wechsler2002,vandenBosch2002,Zhao2009,Ludlow2013,Ludlow2014}.
Here we do not intent to make a comprehensive analysis of the individual halo assmbly tracks, but
provide typical examples of how concentrations and some other basic
properties of individual very large halos -- as opposed to the average -- evolve over
time.

Our main focus is on halos that are located at the upturn or on the
plateau of the concentration-mass relation. Being the most massive,
these halos typically grow very fast at large redshifts. Does it mean,
as it was argued by \citet{Ludlow2012}, that these halos are
substantially out of equilibrium?

For our analysis we use halo tracks selected in the BolshoiP simulation.
Because we are interested in the evolution of very large halos, this
simulation provides a compromise between mass resolution and
halo statistics. 

\begin{figure*}
\centering
\includegraphics[width=0.48\textwidth]{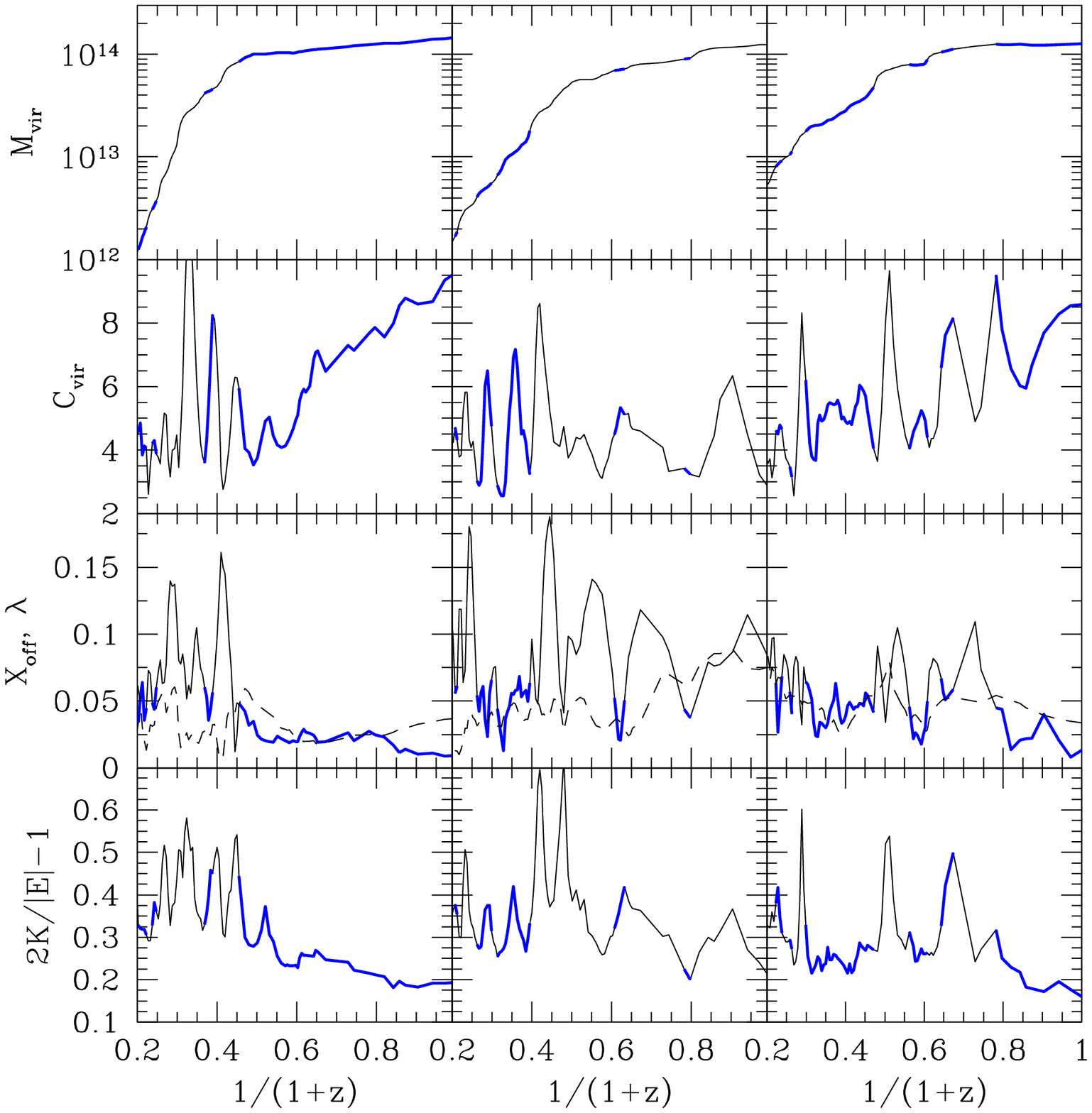}
\includegraphics[width=0.48\textwidth]{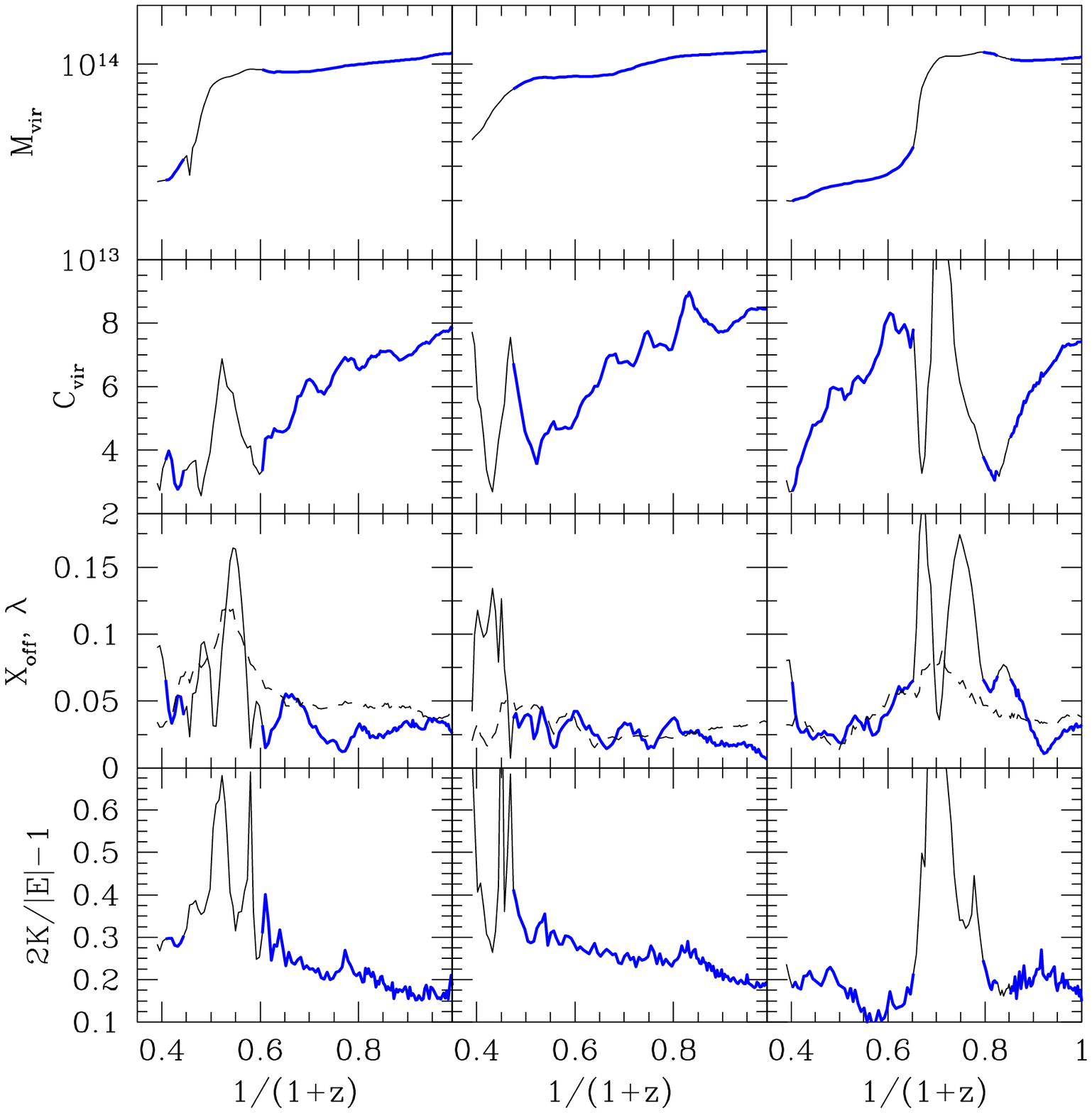}
\caption{Examples of the evolution of virial mass $\Mvir$ ,
  concentration $C_{\rm vir}$, spin parameter $\lambda$ (dashed curves
  in the second from the bottom panels), offset parameter
  $X_{\rm off}$, and virial ratio $2K/|E|-1$ for 6 cluster-size halos
  taken from the BolshoiP simulation. Halos were selected to have
  $\Mvir\approx 10^{14}\Msunh$ and be relaxed at $z=0$. Thick solid (blue)
  parts of the curves indicate that halos are considered to be
  relaxed. Large variations in halo concentration are seen at high
  redshifts when the halo mass increases very quickly. Once the mass
  accretion slows down at low redshifts, halo concentration shows the
  tendency to increase. Major merger events, in the right panels, seen as
  large jumps in mass are followed by temporary increase in halo
  concentration. Most of these major-merger spikes in concentration
  are identified as happening in non-relaxed halos.}
\label{fig:CompareThree}
\end{figure*}

We start with presenting tracks of six halos that have at $z=0$ a mass of
$\Mvir \approx 10^{14}\Msunh$.  The halos were selected to be relaxed
at $z=0$, but no other selection conditions were
used. Figure~\ref{fig:CompareThree} shows the evolution of different
global parameters of these halos.  Large variations in halo
concentration are seen at high redshifts when halo mass increases very
quickly. Once the mass accretion slows down at low redshifts, halo
concentration shows the tendency to increase with time. This behavior
is well known and has been well studied
\citep[e.g.,][]{Bullock2001,Wechsler2002}.

There is also another trend with time: the virial ratio $2K/|E|-1$
tends to decrease with time as the concentration increases. This is in
line with the arguments put forward in
Section~\ref{sec:Relaxed}. Indeed, as the concentration increases,
the effects of the surface pressure become smaller, and the virial ratio
also have a tendency to be smaller.

The three right panels in Figure~\ref{fig:CompareThree} show examples
of major mergers in the regime when the mass accretion slows down at
low redshifts. For example, the first halo on the right experiences
the merging event at $z\approx 1$ (expansion parameter $a=0.5$). Just
as the merging event proceeds, the halo concentration shows a sharp
increase by climbing from $C_{\rm vir}\approx 3$, before the merger, to
the peak value of $C_{\rm vir}\approx 7$. This happens when the
infalling satellite reaches the center of halo, increasing the mass
and, thus, halo concentration before bouncing back. This is
definitely a sign of an out-of-equilibrium event. The jump in
concentration disappears quickly after the onset of the merger. The
same effect during major merger event was found by \citet{Ludlow2012}. Most
of the time, but not always, this jump in concentration is flagged as
happening in a non-relaxed halo.
This behavior of halo concentration is typical for halos at low
redshifts (or at low $\sigma$ peaks): rare major mergers result in
jumps in mass followed by bumps in halo concentration. 

The situation with mass accretion and concentration is more
complicated in the case of fast growing high-$\sigma$ halos. Halos in
the left side of Figure~\ref{fig:CompareThree} indicate that at $z>1$,
when halos quickly increase  mass, their concentration experiences
large fluctuations and overall do not grow with time. 

In order to explore this domain of halo evolution in more detail, we
focus on the most massive halos at $z\sim 2-5$. Specifically, we
study large halos with $\Mvir>2\times 10^{13}\Msunh$ at
$z=2.5$. The vast majority ($\sim 90$\%) of these halos more than double
their mass in a dynamical time, which for these halos is typically equal
to $t_{\rm vir}\approx r_{vir}/V_{\rm vir} = 4.7\times
10^8$yrs.
However, the crossing time of the central region, which defines the
halo concentration, is significantly shorter:
$t_{\rm cross}\approx 2r_s/V_{\rm rms} = 1.2\times 10^8$yrs, where
$r_s= r_{vir}/C$ is the core radius, and $V_{\rm rms}$ is the 3d rms
velocity of dark matter particles in the central region.  This short
central time-scale allows the central halo region to quickly relax and
adjust to ever infalling numerous satellites.

\makeatletter{}

We can approximately split all halos at that redshift and mass range
into ``upturn'' and ``normal'' halos by considering halos with $C>4$
to be in the upturn (see Figure~\ref{fig:conc_vmax}). Overall, about
2/3 of all halos are in the upturn and 1/4 of all halos are relaxed
upturn halos.

\begin{figure*}
\centering
\includegraphics[width=0.48\textwidth]{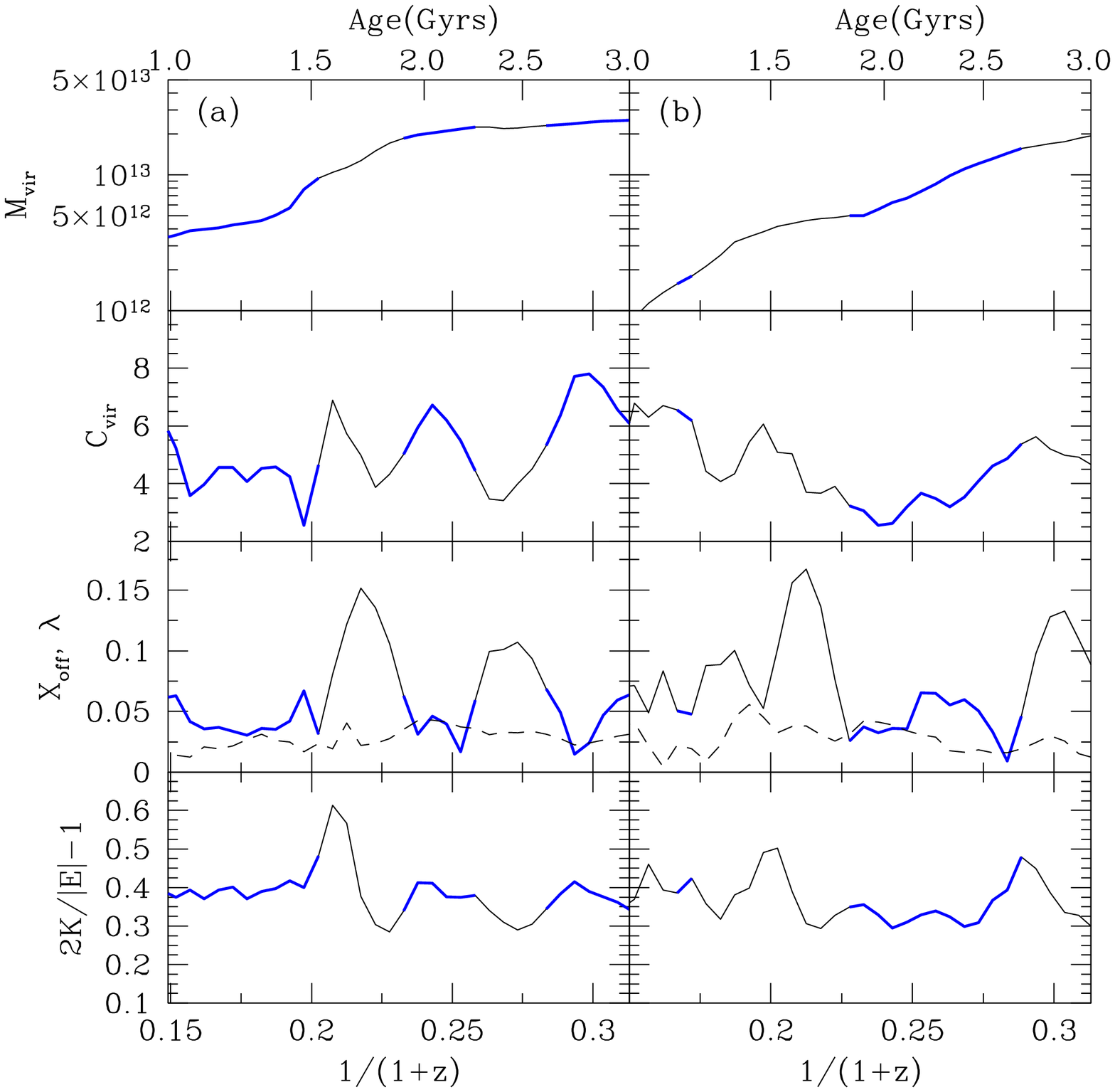}
\includegraphics[width=0.48\textwidth]{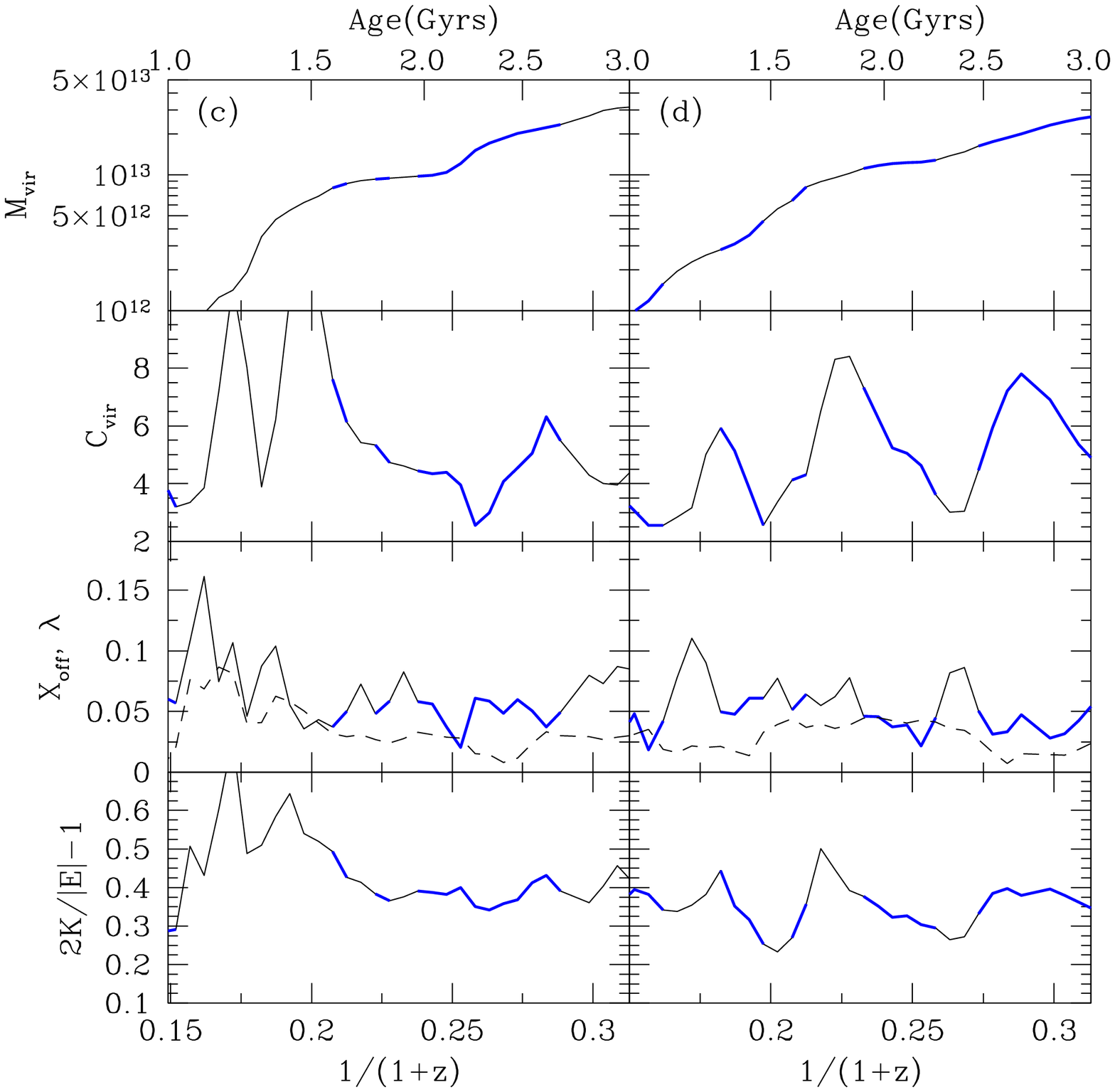}
\caption{Examples of the evolution of parameters for halos identified
  at $z=2.5$ (expansion parameter $a=1/(1+z)= 0.285$) as being relaxed
  and at the upturn of concentration-mass relation ($C>5$).  Notations are the
  same as in Figure~\ref{fig:CompareThree}. Top axes show the age of
  the Universe. All these halos have a very steep growth of mass
  typically increasing mass by a factor of ten over 2~Gyrs ($\sim 4$
  crossing times).  Most of the growth is not related with major-merger
  events and is due to the fast accretion of many small satellite halos. }
\label{fig:CompareUp1}
\end{figure*}

Figure~\ref{fig:CompareUp1} shows the evolution of four upturn
halos. These halos were selected to be relaxed and have large
concentration $C\gsim 5$ at $z=2.5$. The mass accretion history of
these halos shows occasional major merger events in their past, but
there only few of theses events, and most of the mass accretion
happens as a steady and fast mass growth, not major merger. This is
also confirmed by studying masses and positions of largest
subhalos. In all cases the most massive subhalo had just (1-3)\% of
the halo mass at $z=2.5$.  However, there are many of the subhalos and
their total mass is substantial with the first 20 most massive
subhalos providing (10-15)\% of the total halo mass.

For example, halo (a) (left-most halo in Figure~\ref{fig:CompareUp1})
has a major merger event at $t=1.5$~Gyrs, which was followed by a
``normal'' spike in its concentration. At $z=2.5$
($t\approx 2.7$~Gyrs) it had large concentration that is clearly not
related with the early major merger. Indeed, it does not show any
large subhalo moving through its central region that would be
responsible for the large concentration.  Other halos in the figure do
not have signatures of major mergers at $t=1.5-3$~Gyrs.

\makeatletter{}

\begin{figure*}
  \centering
\includegraphics[width=0.48\textwidth]
{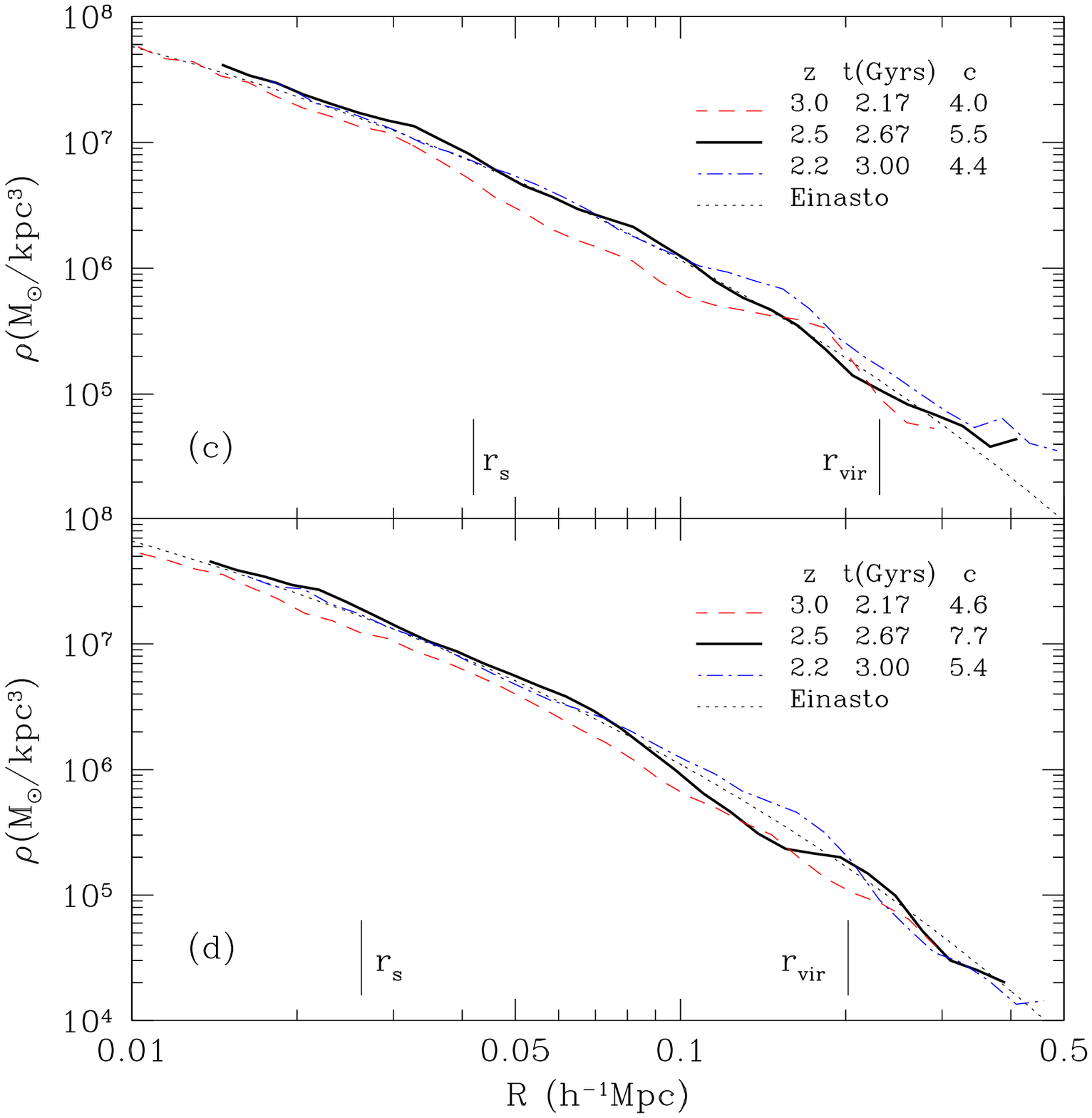}
\includegraphics[width=0.48\textwidth]
{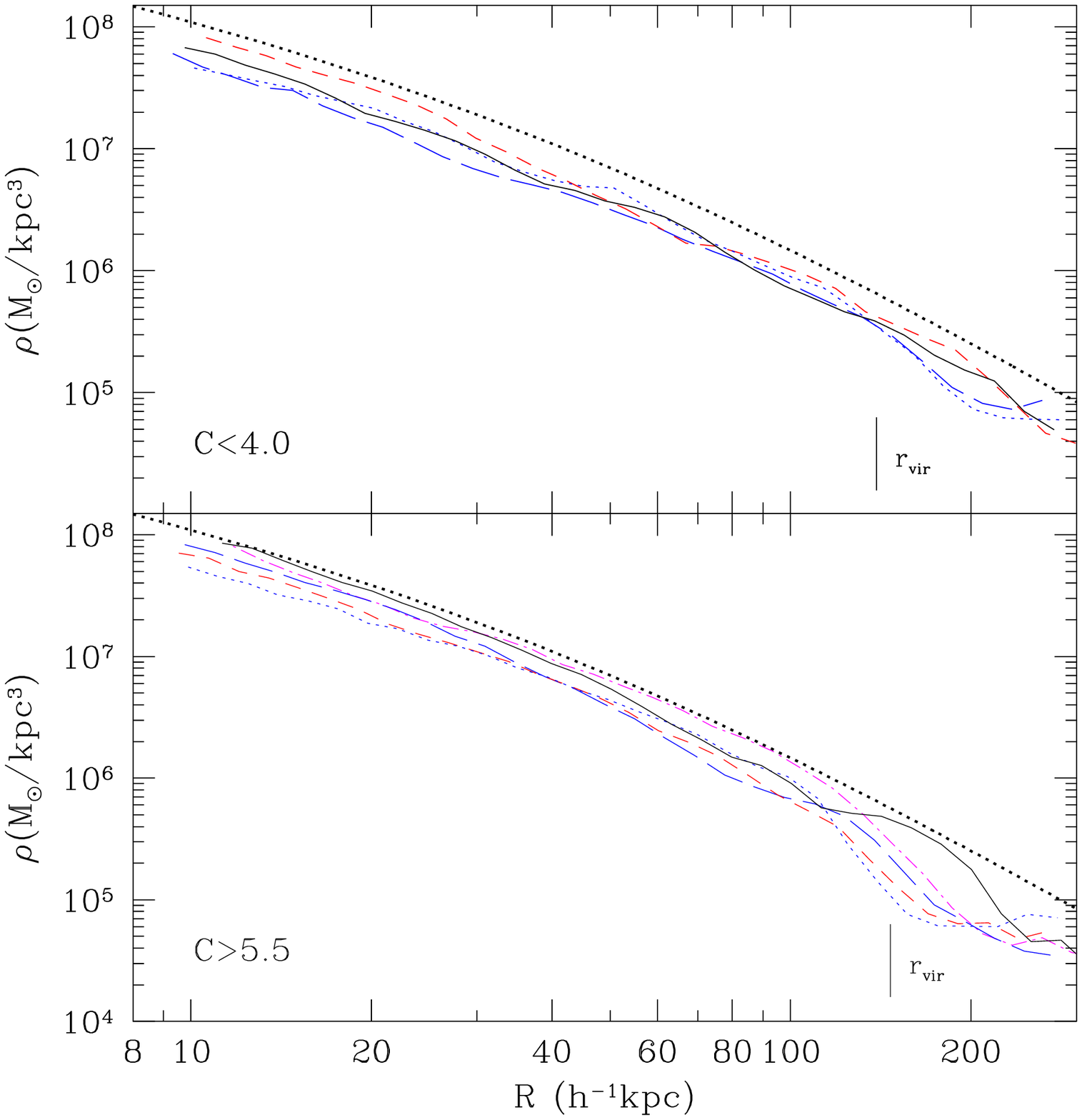}
\caption{Examples of density profiles of individual halos. {\it Left:}
 Evolution of the density profiles for halos {\it d} and {\it c}
  shown in Figure~\ref{fig:CompareUp1}. Both density and radii are given
  in proper (not comoving) units. Halos are selected to be at the
  upturn of the concentration-mass relation at $z=2.5$ (thick solid
  curves). At that moment they have virial mass
  $\Mvir\approx 2\times 10^{13}\Msunh$ and represent $3.5\sigma$
  fluctuations in the density field. Dotted curves show Einasto fits
  with parameter $\alpha =0.29$ (see eq.~(23)) and
  $r_{-2}=R_{vir}/4$. Ages, redshifts, and concentrations are also
  given for each moment presented in the plot.  
  {\it Right:} Examples of density profiles of large
  $(1.2-1.8)\times 10^{13}\Msunh$ relaxed halos at $z=3$. The bottom
  panel shows five halos with large concentration $C>5.5$ that are at
  the upturn part of the halo concentration-mass relation. The top
  panel presents four halos with low concentration $C<4$. Density
  profiles of high concentration halos do not show any signs of large
  out-of-equilibrium interactions or signatures of large recent
  mergers in the central $\sim \Rvir/2$ regions of upturn halos. In
  that respect they are similar to density profiles of low $C$ halos.  There
  are systematic differences in the density profiles of upturn and
  low-$C$ halos with the former being slightly denser in the central
  regions and showing larger curvature of $\rho(R)$ profiles at larger radii.
  For reference dotted curves show the same NFW profiles with $C=3.5$ on both
  panels slightly offset in vertical direction.
   }
\label{fig:DensHalocd}
\end{figure*}

In left panels in figure~\ref{fig:DensHalocd} we show the evolution of
halo profiles for two upturn halos presented in
Figure~\ref{fig:CompareUp1}. There are clear signs of large variations
in the density at large radii comparable to the virial radius, and
there are some variations in the central region too. However, the
evolution of the density in the central $R <R_{\rm vir}/2$ region is
better described as a gradual increase, not a sudden change that one may
expect if the halo was drastically out of equilibrium.

The right two panels shown in Figure~\ref{fig:DensHalocd} give examples of high
concentration upturn halos ($C>5.5$, bottom panel) and low
concentration halos ($C<4$, top panel). To guide the eye, the top
dotted curve in each plot shows the same Einasto profile. High
concentration halos are slightly denser and have larger fluctuations
around the virial radius. They also have systematically different
profile shapes with steeper decline in the outer regions.
Still, there are no drastic differences between high and low
concentration halos.

 \makeatletter{}

These results indicate that major mergers, though important and
happening, do not explain the phenomenon of the upturn halos. What we
find is more consistent with the picture where fast and preferentially
radial infall of numerous satellites results in a quasi-equilibrium
drift of the halo concentration. Because the infalling mass settles at
different radii, the concentration does not increase with time but
experiences large up-and-down variations.

\label{lastpage}
\end{document}